\documentclass[onecolumn,twocolappendix,dvipsnames]{aastex63}

\usepackage{amsmath}
\usepackage{comment}
\usepackage{xcolor}

\usepackage{xcolor}
\definecolor{amethyst}{rgb}{0.6, 0.4, 0.8}

\usepackage{lineno}

\received{June 1, 2019}
\revised{January 10, 2019}
\accepted{November 15, 2022}
\submitjournal{ApJ}

\shorttitle{Solar Flares and Magnetic Helicity in Active Regions}
\shortauthors{Liu et al.}

\graphicspath{{./}{figures/}}

\begin{document}

\title{Changes of Magnetic Energy and Helicity in Solar Active Regions from Major Flares}

\correspondingauthor{Yang Liu}
\email{yliu@sun.stanford.edu}

\author{Yang Liu}
\affiliation{W. W. Hansen Experimental Physics Laboratory, Stanford University, Stanford, CA 94305-4085, USA}

\author{Brian T. Welsch}
\affiliation{Natural \& Applied Sciences,
University of Wisconsin-Green Bay, 2420 Nicolet Drive, Green Bay, WI 54311, USA}

\author{Gherardo Valori}
\affiliation{Max-Planck-Institut f{\"u}r Sonnensystemforschung, Justus-von-Liebig-Weg 3, 37077, G{\"o}ttingen,Germany}

\author{Manolis K.Georgoulis}
\affiliation{Research Center for Astronomy and Applied Mathematics of the Academy of Athens, 11527 Athens, Greece}

\author{Yang Guo}
\affiliation{School of Astronomy and Space Science, Nanjing University, Nanjing 210023, China
}

\author{Etienne Pariat}
\affiliation{Sorbonne Universit\'e, \'Ecole polytechnique, Institut Polytechnique de Paris, Universit\'e Paris Saclay, Observatoire de Paris-PSL, CNRS, Laboratoire de Physique des Plasmas (LPP), 75005 Paris, France}

\author{Sung-Hong Park}
\affiliation{W. W. Hansen Experimental Physics Laboratory, Stanford University, Stanford, CA 94305-4085, USA}

\author{Julia K. Thalmann}
\affiliation{University of Graz, Institute of Physics/IGAM, Graz, Austria}
\begin{abstract}

Magnetic free energy powers solar flares and coronal mass ejections (CMEs), and the buildup of magnetic helicity might play a role in the development of unstable structures that subsequently erupt.  To better understand the roles of energy and helicity in large flares and eruptions, we have characterized the evolution of magnetic energy and helicity associated with 21 X-class flares from 2010 to 2017.  Our sample includes both confined and eruptive events, with 6 and 15 in each category, respectively. Using HMI vector magnetic field observations from several hours before to several hours after each event, we employ (a) the Differential Affine Velocity Estimator for Vector Magnetograms (DAVE4VM) to determine the photospheric fluxes of energy and helicity, and (b) non-linear force-free field (NLFFF) extrapolations to estimate the coronal content of energy and helicity in source-region fields.

Using Superposed Epoch analysis (SPE), we find, on average: (1) decreases in both magnetic energy and helicity, in both photospheric fluxes and coronal content, that persist for a few hours after eruptions, but no clear changes, notably in relative helicity, for confined events; (2) significant increases in the twist of photospheric fields in eruptive events, with twist uncertainties too large in confined events to constrain twist changes (and lower overall twist in confined events); and (3) on longer time scales (event time +12 hours), replenishment of free magnetic energy and helicity content to near pre-event levels for eruptive events. For eruptive events, magnetic helicity and free energy in coronal models clearly decrease after flares, with the amounts of decrease proportional to each region's pre-flare content.

\end{abstract}



\section{Introduction}
\label{s:intro}

Solar flares and coronal mass ejections (CMEs) are driven by the release of free magnetic energy stored in coronal electric currents.   Beyond the basic outline of this ``storage and release'' paradigm, however, key aspects of how magnetic energy builds up in coronal magnetic fields and the processes that trigger its impulsive release in flares and CMEs are poorly understood.  Improved knowledge of these build-up and triggering processes, however, could benefit efforts to forecast these drivers of space weather disturbances \citep{Baker1998}, suggest possible physical processes at work in stellar flares and ejections, and aid understanding of magnetic energy release from plasmas generally.  

Electric currents in the corona are the source of free magnetic energy --- i.e., energy above the current-free magnetic field
(also known as a potential field) with the same normal field below the base of the corona, in the solar photosphere. This potential field is unique, given conditions of isolated magnetic structures and extension to infinity \citep[i.e.,][]{sakurai89}, and possesses the minimum possible magnetic energy for a given set of boundary conditions. Using this potential field as a ``reference field,'' the coronal field's relative magnetic helicity is a gauge-invariant measure of the coronal field's connectivity and complexity (i.e., twist, writhe, and linkage - for a brief review of magnetic helicity, see, for instance, \citealt{Berger1998b}.)  A nonzero relative magnetic helicity thus signifies the presence of non-potential, current-carrying magnetic structures, such as flux systems with twist (helical winding of field lines around each other within a flux system) or writhe (helical deformation of the axis of a flux system), and sheared fields (i.e., horizontal fields that point along a polarity inversion line [PIL] of the normal magnetic field, rather than across the PIL as a generic potential field would). 
For simplicity, we use the term helicity to refer to relative helicity throughout the paper, unless otherwise specified.

The availability of routine measurements of the photospheric magnetic field vector in active regions (ARs) from the Helioseismic and Magnetic Imager (HMI; \citealt{Scherrer2012,Hoeksema2014}) aboard the Solar Dynamics Observatory (SDO; \citealt{Pesnell2012}) has enabled estimating the time evolution of (i) fluxes of magnetic energy (the Poynting flux) and helicity across the photosphere (e.g., \citealt{Liu2012d, Kazachenko2015, Park2020, Liokati2022}) and (ii) magnetic energy and helicity in coronal volumes where, typically, the magnetic field is extrapolated by assuming that it is force-free (e.g., \citealt{Wiegelmann2012b, Thalmann2016, moraitis2019}).

Variations in the photospheric fluxes or coronal content of magnetic energy and helicity associated with the occurrence of flares and CMEs have been investigated in several studies. Particular attention has been focused on AR magnetic properties that differ between eruptive and confined flares. \citet{tziotziou2012,tziotziou2014} found that the eruptive events appear fairly segregated from confined events in both free energy and relative helicity. When the helicity is broken down into two components --- the magnetic helicity of the non-potential component of the magnetic field (the current-carrying field), $H_j$, and the mutual helicity between potential field and current-carrying field, $H_{pj}$ \citep{Berger2003,pariat2017,linan2018} --- changes in the ratio between these components could mark the trigger of an eruptive flare \citep[e.g.,][]{pariat2017,zuccarello2018}.  This ratio effectively divides the eruptive and confined events well \citep{thalmann2019,gupta2021}. Thus far, comparisons between the two types of helicity estimation have been limited to individual case studies \citep[see, e.g.,][]{Thalmann2021}.

In this study, we systematically analyze and compare the photospheric fluxes and coronal content of magnetic helicity and energy in a set of 13 ARs that produced 21 X-class flares during solar cycle 24. Unlike previous studies, we focus specifically on changes of these quantities associated with these major eruptive and confined flares. In addition, we seek and discuss possible differences in the behavior of time series for these two flare categories.

In the following Section \ref{s:methoddata}, we describe our event sample and methods for characterizing the evolving non-potentiality of these flare-productive active regions.  In Section \ref{s:result}, we present quantitative results of our analyses.   In Section \ref{s:conclusion}, we recap our most significant findings, consider their implications, and discuss future research directions of research.


\section{Data and Methodology} 
\label{s:methoddata}

We first describe our procedures for event selection, and then describe the observational data for our event sample.  We then describe our procedures for analyzing magnetic energy and helicity derived from these observations.

\subsection{Analysed Data Set}
\label{ss:sampledata}

\subsubsection{Event Sample}
\label{sss:sample}

We evaluated all X-class flares in the time period, ranging between 2010 to 2017, for which HMI observations are available. We found 28 ARs that produced X-class flares during this period, out of which we selected 11 ARs that produced an event associated with a CME (i.e., an eruptive flare) and two ARs that produced confined events. The remaining ARs only produced X-class events that occurred relatively close to the limb ($>50^{\circ}$ from the central meridian) or were located in relatively weak fields at the 
periphery of an evolving AR (such as the 2014 January 17 X1.2 flare in AR 11944). Because the HMI vector data are less reliable in such cases, we excluded these events from our sample. 
We also excluded the 2017 September 6 X2.2 flare in AR 12673 because it is unclear if this flare is eruptive or confined. This flare was reported to be an eruptive event by \citet{mitra2018}, who relied on  the Solar Eruptive Event Detection System (SEEDS) CME catalog\footnote{\url{http://spaceweather.gmu.edu/seeds/monthly.php?a=2017&b=09}}: it determined that a CME occurred near that time, with an angular width of 44$^{\circ}$ and a speed of 279 km s$^{-1}$. 
However, \citet{liu2018} reported this flare as a confined event because they did not observe separation of flare ribbons or clear CME signatures in He II 304\AA~ images.  They also reported that no dimmings were visible, though we observed minor dimmings in our review of AIA images.
The uncertainty regarding the eruptive nature of this event precluded us from placing it in either the confined or the eruptive category.

Overall, our sample includes 21 X-class flares from 13 different ARs. Six flares from two ARs in the sample are confined. All events are listed in Table \ref{table:xflares}. Because the two X-class flares in AR 11429 occurred about 20 minutes apart, they are deemed to be one event in our analysis. 

\begin{center}
\begin{table}[htp]
\caption{
List of X-class flares in 13 ARs in our sample, all of which are within $50^\circ$ from the central meridian and occurred after May 2010. Six flares in the sample of 21 events are confined, i.e., they have no CMEs associated to them.
}
\vspace{0.3cm}
\scriptsize
\begin{tabular}{llllllll}

\hline
Flare ID & Flare Start & Peak & End & Flare & CME & AR & Position \\
& & & & & & \\

\hline

& & & & & & & \\

1 & 2011-02-15T01:44:00 & 2011-02-15T01:56:00 & 2011-02-15T02:06:00 & X2.2 & YES & 11158 & S21W28\\
2 & 2011-03-09T23:13:00 & 2011-03-09T23:23:00 & 2011-03-09T23:29:00 & X1.5 & NO  & 11166 & N11E13\\
3 & 2011-09-06T22:12:00 & 2011-09-06T22:20:00 & 2011-09-06T22:24:00 & X2.1 & YES & 11283 & N14W18\\
4 & 2011-09-07T22:32:00 & 2011-09-07T22:38:00 & 2011-09-07T22:44:00 & X1.8 & YES & 11283 & N14W18\\
5 & 2012-03-07T00:02:00 & 2012-03-07T00:24:00 & 2012-03-07T00:40:00 & X5.4 & YES & 11429 & N17E27\\
6 & 2012-03-07T01:05:00 & 2012-03-07T01:14:00 & 2012-03-07T01:23:00 & X1.3 & YES & 11429 & N17E26\\
7 & 2012-07-12T15:37:00 & 2012-07-12T16:49:00 & 2012-07-12T17:30:00 & X1.4 & YES & 11520 & S15W01\\
8 & 2013-11-05T22:07:00 & 2013-11-05T22:12:00 & 2013-11-05T22:15:00 & X3.3 & YES & 11890 & S09E36\\
9 & 2013-11-08T04:20:00 & 2013-11-08T04:26:00 & 2013-11-08T04:29:00 & X1.1 & YES & 11890 & S09W20\\
10 & 2013-11-10T05:08:00 & 2013-11-10T05:14:00 & 2013-11-10T05:18:00 & X1.1 & YES & 11890 & S09W30\\
11 & 2014-03-29T17:35:00 & 2014-03-29T17:48:00 & 2014-03-29T17:54:00 & X1.0 & YES & 12017 & N11W32\\
12 & 2014-09-10T17:21:00 & 2014-09-10T17:45:00 & 2014-09-10T18:20:00 & X1.6 & YES & 12158 & N12E29\\
13 & 2014-10-22T14:02:00 & 2014-10-22T14:28:00 & 2014-10-22T14:50:00 & X1.6 & NO  & 12192 & S12E20\\
14 & 2014-10-24T21:07:00 & 2014-10-24T21:41:00 & 2014-10-24T22:13:00 & X3.1 & NO  & 12192 & S12W15\\
15 & 2014-10-25T16:55:00 & 2014-10-25T17:08:00 & 2014-10-25T18:11:00 & X1.0 & NO  & 12192 & S12W28\\
16 & 2014-10-26T10:04:00 & 2014-10-26T10:56:00 & 2014-10-26T11:18:00 & X2.0 & NO  & 12192 & S12W35\\
17 & 2014-10-27T14:12:00 & 2014-10-27T14:47:00 & 2014-10-27T15:09:00 & X2.0 & NO  & 12192 & S12W45\\
18 & 2014-11-07T16:53:00 & 2014-11-07T17:26:00 & 2014-11-07T17:34:00 & X1.6 & YES & 12205 & N15E33\\
19 & 2014-12-20T00:11:00 & 2014-12-20T00:28:00 & 2014-12-20T00:55:00 & X1.8 & YES & 12242 & S18W29\\
20 & 2015-03-11T16:11:00 & 2015-03-11T16:22:00 & 2015-03-11T16:29:00 & X2.1 & YES & 12297 & S16E13\\
21 & 2017-09-06T11:53:00 & 2017-09-06T12:02:00 & 2017-09-06T12:10:00 & X9.3 & YES & 12673 & S09W38\\
& & & & & & &\\
\hline
\end{tabular}
\label{table:xflares}
\end{table}
\end{center}

\subsubsection{Data}
\label{sss:data}

We use vector magnetic field data taken by HMI, a filtergraph instrument with full-disk coverage over 4096$\times$4096 pixels. Its spatial resolution is about 1", with a 0.5" pixel size.  The spectral line observed is the Fe{\sc i} 6173\AA~ absorption line formed in the photosphere \citep{norton2006}.  There are two CCD cameras in the instrument, the ``front camera'' and the ``side camera.''  The front camera acquires filtergrams at 6 wavelengths along the line Fe{\sc i} 6173\AA~ in two polarization states with 3.75 seconds between the images. The width of the filter profile is 76 m\AA.  It takes 45 seconds to acquire a set of 12 filtergrams. This set of data is used to derive Dopplergrams and line-of-sight magnetograms. 
The side camera is dedicated to measuring the vector magnetic field. It takes 135 seconds to obtain the filtergrams in 6 polarization states at 6 wavelength positions. After April 2016, the observation mode changed: the side camera now only takes linear polarization states at 6 wavelengths in 90 seconds, and measurements from both cameras are combined. The Stokes parameters [I, Q, U, V] are computed from these measurements.

In order to suppress the p-modes and increase the signal-to-noise ratio, usually the Stokes parameters are derived from filtergrams averaged over 1215 seconds using a cosine-apodized, moving-boxcar weighting function with a FWHM of 720 seconds.  Averages are computed at a 720-second cadence.  They are then inverted to produce the vector magnetic field using the Very Fast Inversion of the Stokes Vector (VFISV) algorithm \citep{borrero2011,centeno2014}. The 180$^\circ$ degree ambiguity in the transverse-field azimuth is resolved based on the ``minimum energy'' algorithm \citep{metcalf1994,leka2009}. Photospheric patches containing  ARs are automatically identified and bounded by a feature recognition model \citep{turmon2010}, and the disambiguated vector magnetic field data of ARs are deprojected to heliographic coordinates \citep{bobra2014}. Here we use the Lambert (cylindrical equal area) projection method, centered on each region, for the remapping.

\begin{figure}[htp]
\centering
\includegraphics[width=0.50\textwidth]{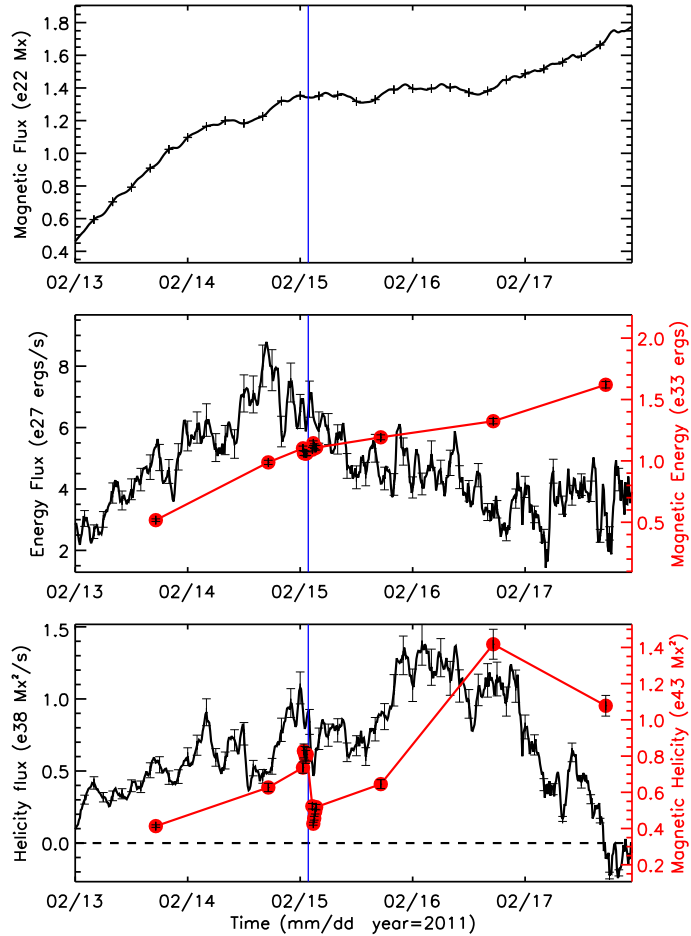}
\vspace{0.5cm}

\caption
{
Temporal profiles of magnetic flux (top, total unsigned flux divided by 2), energy flux with total energy (middle), and helicity flux and total helicity (bottom) in AR 11158 from 2011 February 13--17. Right axes on middle and bottom panels show magnetic energy and helicity, respectively, calculated from NLFFF models (see Section \ref{ss:method} in the text). Vertical blue lines denote the occurrence of an X-class flare that started at 01:44 UT, 2011 February 15. One-$\sigma$ uncertainties are shown at representative data points. }
\label{fig:example11158}
\end{figure}

\subsection{Analysis Methodology}
\label{ss:method}
\noindent

For each AR in this study, we used the time series of its photospheric observations to estimate time series of relative helicity and magnetic energy injection rates through the solar surface, as well as the relative helicity and energy in extrapolated coronal fields. 
We also calculate time series of magnetic twist in each AR. We use force-free field parameter, $\alpha$, to represent the magnetic twist in AR (See details in Section \ref{ss:twist change}).

The relative magnetic helicity in a volume $V$ can be expressed by \citep{finn1985},
\begin{equation}
H = \displaystyle \int_{V} (\mathbf{A} + \mathbf{A}_p) \cdot (\mathbf{B} - \mathbf{B}_p) \, dV ~, \label{eq:RelH}
\end{equation}
where $\mathbf{B}$ and $\mathbf{B_p}$ are the magnetic field and the potential (reference) field, respectively, and $\mathbf{A}$ and $\mathbf{A_p}$ are their respective vector potentials. 
Details of helicity calculation are given by \citet{valori2012}. Specifically in this computation, the vector potentials $\mathbf{A}$ and $\mathbf{A_p}$ are computed assuming that the vertical component of the vector potentials vanish everywhere in the considered volume $V$ (DeVore gauge, see in particular Equations~11, 24, and 25 in \citet{valori2012}). Note that this is different from the gauge used for the helicity flux calculation, as described 
below. 
The volume helicity is gauge independent, so the choice of gauge does not impact the helicity calculation. The difference between gauges was also tested numerically \citep{valori2016}, and no significant difference was found. The method chosen here to compute the volume helicity is computationally 
faster.

Relative helicity can be broken down into two gauge-invariant components, the helicity of the current-carrying field,
$H_j$, and the mutual helicity between potential field and current-carrying field, $H_{pj}$, expressed as \citep{Berger2003,pariat2017},
\begin{equation}
H = H_j + H_{pj} ~,
\end{equation}
where 
\begin{equation}
H_j \equiv \displaystyle \int_{V} (\mathbf{A} - \mathbf{A}_p)\cdot (\mathbf{B} - \mathbf{B}_p) dV ~,
\label{eq:hj}
\end{equation}
and 
\begin{equation}
H_{pj} \equiv 2\times \displaystyle \int_{V}\mathbf{A}_p\cdot (\mathbf{B} - \mathbf{B}_p) dV ~.
\end{equation}

The magnetic energy in a volume $V$ is computed as
\begin{equation}
E = \displaystyle \int_{V} \frac{B^2}{8\pi} \, dV ~.
\end{equation}

The coronal magnetic field for each AR can be extrapolated from the observed vector magnetic field at the photosphere by assuming that the Lorentz force vanishes in the coronal field. We use an optimization-based non-linear force-free field (NLFFF) algorithm \citep{wiegelmann2004} to extrapolate the coronal field in a Cartesian volume above a given photospheric magnetogram. A preprocessing step alters the magnetogram field to remove most of the net force and torque from the photospheric data, thereby making the boundary more consistent with the force-free assumption \citep{wielgelmann2006}.

The flux of magnetic helicity across a surface {\it S} is expressed by \citep{berger1984,Chae2001},
\begin{equation}
\frac{dH}{dt}\Bigg| _{S} = 2\displaystyle \int_{S} (\mathbf{A}_p\cdot\mathbf{B}_t)V_{\perp n} \, dS 
- 2\displaystyle \int_{S} (\mathbf{A}_p\cdot\mathbf{V}_{\perp t})B_n \, dS,  \label{eq:hflux}
\end{equation}
where: 
{\bf B}$_t$ and B$_n$ denote the observed tangential and normal magnetic fields, respectively; and {\bf V}$_{\perp t}$ and V$_{\perp n}$ are the tangential and normal components of ${\mathbf{V}_\perp}$, the velocity perpendicular to the magnetic field. The integral is done over the surface imaged in the magnetogram.   The vector potential {\bf A}$_p$ for the potential field on the photosphere is uniquely determined by the observed
photospheric normal magnetic field and Coulomb gauge with the following equations \citep{berger1997,berger2000},
\begin{equation}
\mathbf{\nabla} \times \mathbf{A}_p\cdot \hat{\mathbf{n}} = B_n, \mathbf{\nabla} \cdot \mathbf{A}_p = 0, \mathbf{A}_p\cdot \hat{\mathbf{n}} = 0.
\end{equation}

Similarly, the magnetic energy (Poynting) flux can be calculated by
\begin{equation} 
\frac{dE}{dt}\Bigg| _{S} = \frac{1}{4\pi}\displaystyle \int_{s} {B}_t^2 V_{\perp n} \, dS 
- \frac{1}{4\pi}\displaystyle \int_{S} (\mathbf{B}_t\cdot\mathbf{V}_{\perp t})B_n \, dS. 
\end{equation}
The vector velocity field in the photosphere is derived from the Differential Affine Velocity Estimator for Vector Magnetograms \citep[DAVE4VM;][]{schuck2008}, applied to the time-series of reprojected, co-registered vector magnetic field data. The window size used in DAVE4VM is 19 pixels, which is about 9.5 arcsec.  DAVE4VM derives velocities at time $t_{i+1/2}$ from (i) the difference $\Delta \mathbf{B}$ between HMI's measurements of $\mathbf{B}$ at times $t_i$ and $t_{i+1}$ and (ii) the average field from these two measurements.  Nominal cadence between HMI measurements is 720 s, so the nominal $\Delta t$ between flow estimates is also 720 s.

This window size was chosen based on a combined evaluation of how well the recovered flows satisfied the ideal MHD induction equation's vertical component, and how stable the calculated helicity and energy fluxes were, as described by \citet{schuck2008} and \citet{liuschuck2012}. Tests of DAVE4VM with simulation data by \citet{schuck2008} showed that the helicity and energy fluxes became insensitive 
if the chosen window size exceeded a certain value, and that 
the helicity flux from DAVE4VM velocities was about 94\% of the true flux. With HMI data for AR 11158, \citet{Lumme2019} showed the {\em shapes} of the time series of energy and helicity fluxes from different choices of $\Delta t$ and window sizes were similar, though they differed in magnitude. 
This suggests that different choices of window size could still show flare-related changes in helicity and energy fluxes, but the sizes of the changes found might differ.

We remark that the cumulative (time-integrated) fluxes of energy and helicity may not be identical with estimates of the coronal content for various physical reasons. First, the coronal models contain only a finite subdomain of the corona, therefore, discrepancies could occur due to fields that close outside coronal models' domains. For energy, differences between integrated Poynting fluxes and coronal models' energies could be due to losses from coronal heating and flares smaller than the X-class events that are not our focus. In contrast, for helicity, which is expected to be approximately conserved
even when magnetic reconnection occurs in plasmas with a low magnetic Reynolds number  \citep[][]{Berger1984c,Berger1999b}, CMEs or smaller-scale ejections could bodily remove helicity from the coronal volume. Finally, the helicity flux in Equation (\ref{eq:hflux}) is intrinsically derived assuming an infinite plane, while the relative helicity defined in Equation (\ref{eq:RelH}) supposes a finite volume. This can lead to discrepancies between the two estimates. Comparisons between estimates of energy and helicity from fluxes versus extrapolations are underway in a separate study (Pariat et al. 2022, in preparation). 

\section{Results}
\label{s:result}

As an example output of our calculation methods, Figure \ref{fig:example11158} shows temporal profiles of magnetic flux (top), magnetic energy and Poynting flux (middle), and magnetic helicity and helicity injection rate (bottom) for AR 11158 from 13--18 February 2011. The blue vertical lines denote the occurrence of the X2.2 flare that was associated with a CME. The magnetic flux plotted is 
taken as half of the
total unsigned flux, which is computed 
by summing the absolute value of radial field over the AR. The pixels with total field strength greater than 300 G, which is about three times the estimated 1-$\sigma$ transverse-field noise level for HMI vector magnetograms \citep{Hoeksema2014}, are included in these calculations.

The 1-$\sigma$ error bars for each quantity are given at representative points. To estimate the uncertainties in helicity and energy fluxes, we adopt the results in \citet{Liu2012d} that are estimated by employing the Monte Carlo method: for each run, the magnetograms are perturbed by their estimated errors, velocities are derived from the perturbed magnetograms, and the fluxes are computed from the perturbed fields and derived velocities. This uncertainty estimate does not include errors in the velocity inversion from DAVE4VM. As a measure of the uncertainty in helicity and energy estimates for this AR, we overplotted the standard deviations of helicity and energy calculations from 5 consecutive NLFFF models, derived from vector magnetograms at 720-second cadence. 
Assuming that the five consecutive observations (spanning 48 minutes) are repeated measurements of the same quantity, their standard deviation characterizes measurement uncertainty.

For NLFFF models, detailed discussions and careful investigations of the sources of uncertainty in calculating helicity in ARs have been undertaken, including uncertainties such as solenoidality \citep{2019ApJ...880L...6T} parameter settings \citep{2020A&A...643A.153T}, methods to compute helicity \citep{thalmann2019} and spatial sampling of modeled NLFFF data \citep{2022A&A...662A...3T}.

The increase in magnetic flux with time (black curve in top panel of Figure \ref{fig:example11158}) indicates that this is an developing AR, with new flux emerging. This has been noted by previous studies, too \citep[i.e.,][]{Tziotziou2013}. While the coronal magnetic energy does not change significantly after the flare (red curve in the middle panel), the coronal helicity decreases beyond error bars (red curve in the bottom panel). Note that the magnetic energy in the AR includes both potential and free energy. The helicity injection rate also decreases after the event, per the black curve in the bottom panel.

It is essential to combine information from multiple events when considering of flare-associated changes in $E$, $dE/dt$, $H$, and $dH/dt$.  Changes in an individual case could lead to misleading inferences.  For instance, the changes in coronal helicity like that in AR 11158 ($\sim 2 - 3 \times 10^{42}\;Mx^2$ in Figure \ref{fig:example11158}) might be typical following eruptions, or might arise from some unique property of this region.  
Figure \ref{fig:example11158} also demonstrates that the fluxes of energy and helicity regularly exhibit substantial fluctuations. Consequently, any change in these fluxes, even if statistically significant, could be incidental and not necessarily related to a given flare or eruption.

To synthesize information from all events in our sample, we employ superposed epoch analysis (SPE; e.g., \citealt{Chree1913, Wilcox1967, singh2006, Mason2010}).  This approach can highlight variations in time series of a given variable (e.g., $dH/dt$) that are linked with occurrence of some event (e.g., flares).  In SPE, the events' epochs are superposed: time series of the variable of interest are summed after each series is shifted in time to align all event times (the ``key times''). In this summation, random variations in the variable of interest should cancel, but recurring variations will be reinforced.  Below, we present results of SPE applied to our time series of $E$, $dE/dt$, $H$, $dH/dt$, and twist for our event sample.

\subsection{Flare-induced Changes in Helicity and Energy Fluxes}
\label{ss:changeofHEflux}

Because an AR's helicity, $H$, scales as the square of the magnetic flux, $\Phi$, in the AR, combining events from different-sized ARs for SPE will benefit from rescaling their helicities. 
Similarly, larger ARs will, all else equal, have larger Poynting fluxes.  For $dH/dt$ and $dE/dt$, we can simply normalize each time series in the time interval around each flare by each AR's maximum value during the interval considered.  We also take the absolute value of $dH/dt$, to combine unsigned helicity fluxes. 

\begin{figure}[ht]
\centering
\includegraphics[width=0.50\textwidth]{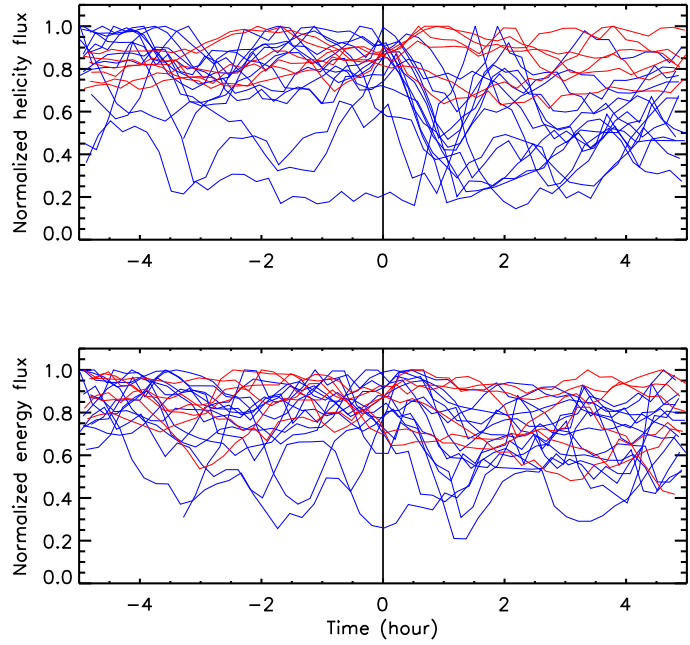}
\vspace{0.5cm}
\caption
{
Normalized, unsigned helicity flux (top) and energy flux (bottom) for all the X-class flares listed in Table \ref{table:xflares}. Each AR's time series is shifted to align the start time of its flare to hour zero, which is marked with a vertical black line in each panel. The curves span 10 hours: 5 hours before and after each region's flare time. Blue curves correspond to eruptive events; red to confined events. In order to better show evolution, a one-hour running average has been applied to each curve.
}
\label{fig:hfluxeflux}
\end{figure}

Figure \ref{fig:hfluxeflux} shows normalized plots of helicity flux $|dH/dt|$ (top panel) and energy flux $dE/dt$ (bottom panel) for both eruptive (blue curves) and confined (red curves) flares within $\pm$5 hours of each flare, with all flare times co-aligned. For eruptive flares, a pattern of decreases shortly after flares can be discerned in both fluxes, though it is more clear in the $|dH/dt|$ plot (top panel). Confined events do not exhibit obvious changes near the flare times. This suggests that eruptions, unlike confined flares, are able to impact the injection process, inhibiting the injection of energy and helicity. Decreases in helicity flux after flares has been reported before \citep{wangrui2016,bi2018}.
 
Two physical scenarios might explain decreases in  energy and helicity flux following eruptive flares: submergence of non-potential / helical magnetic flux associated with the magnetic restructuring occuring in eruptions; and the downward propagation of helicity (e.g., twist propagation) along fields that thread the photosphere.  
Submergence might occur as fields near the flaring PIL that shallowly arch over the PIL become more horizontal immediately after the eruptive flare.  Several observations of strong flares causing fields to become more horizontal have been reported (e.g., \citealt{Hudson2008, Wang2010, Barczynski19}).  Because radial flux only submerges (or emerges) along PILs, areas with negative Poynting flux should be near PILs if flare-associated submergence is responsible for changes in energy fluxes. 

\begin{figure}[ht]
\centering
\includegraphics[width=0.50\textwidth]{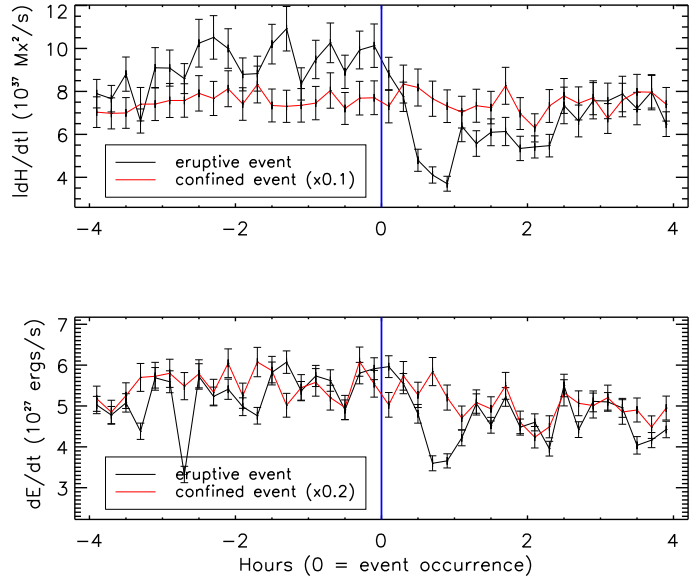}
\vspace{0.5cm}

\caption
{
Superposed Epoch Analysis (SPE) plots for unsigned helicity flux (top) and energy flux (bottom) for all the events. The fluxes are not smoothed. The black curves are averages from all eruptive events; red curves are averages from all confined events. To better compare the results for the two types of the events, the SPE curves for confined events are scaled by multiplying by 0.1 for helicity flux and 0.2 for energy flux.  
(The much larger values for confined events occur because most of these events occur in AR 12192, which contained an unusually large amount of magnetic flux, and both helicity and energy fluxes tend to be larger in larger-flux active regions.) Vertical blue lines denote the flare onset time. One-$\sigma$ uncertainties by propagating individual uncertainties are overplotted.
}
\label{fig:spe4ehflux}
\end{figure}

To better show the general trend of the evolution of helicity and energy fluxes for our events, we average the superposed, not normalized fluxes for the eruptive events, and, separately, for the confined events.  Figure \ref{fig:spe4ehflux} shows the averaged unsigned helicity flux (top) and energy flux (bottom) for  eruptive (black) and confined (red) events.
Uncertainties for each region were propagated in this averaging. For the eruptive events, both helicity and energy fluxes decrease substantially after the events, followed by a 
return toward pre-event levels during the first hours after the event. The onset of change in the helicity flux here precedes the onset of change in the energy flux by roughly one 720-s step.  Given our cadence and measurement uncertainties, however, this difference in onsets is not observationally significant.  
Nevertheless, we note that changes in helicity slightly preceded changes in magnetic energy in the eruption simulations of \citet{Pariat2015a}. A follow-up study with larger sample size and higher-cadence data (e.g., the 135-s HMI magnetograms discussed by \citealt{sun2017highres}) might find this time difference to be significant. For the confined events, no significant changes in the helicity and energy injection rates can be seen.

\begin{figure}[ht]
\centering
\includegraphics[width=0.50\textwidth]{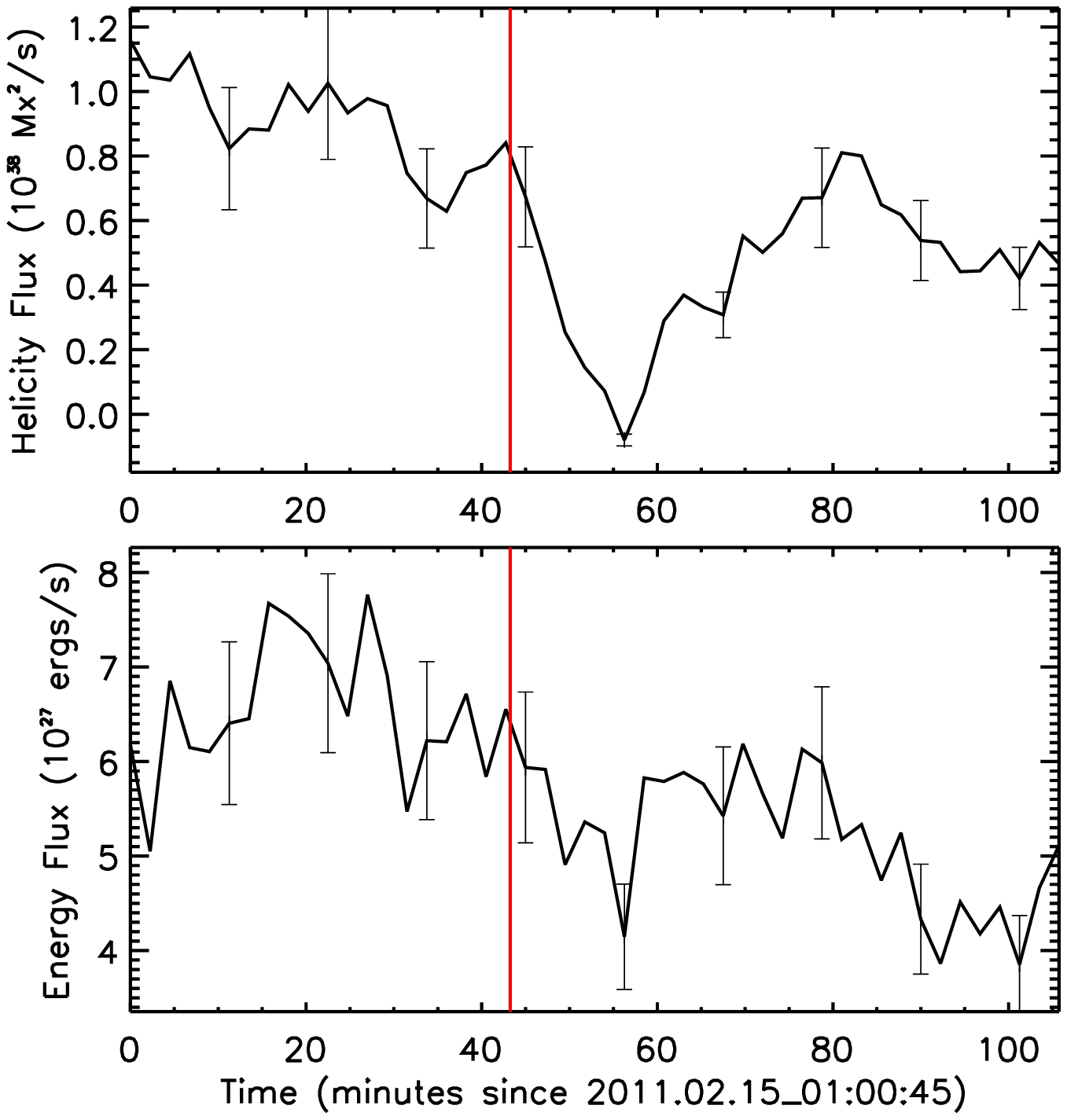}
\vspace{0.5cm}

\caption
{
Temporal profiles of helicity (top) and energy (bottom) flux for 135-second-cadence vector magnetograms of AR 11158. No running average was applied. Bad pixels due to the X-class flare are corrected using the procedure suggested by \citet{sun2017highres}. The red vertical line denotes the flare start time; 1-$\sigma$ uncertainties are given at representative data points.
}
\label{hflux135s}
\end{figure}

It is plausible that the magnitudes of the flare-related changes in helicity and energy fluxes that we find would differ if a different tracking method were used to estimate these fluxes.  
For instance, comparing cumulative helicity and energy fluxes for AR 11158 from various studies, \citet{Kazachenko2015} reported differences of as much as $\sim$30\% and $\sim$25\%, respectively, between DAVE4VM and other methods; see their Table 2.
By comparison, the drops in helicity and energy fluxes in Figure \ref{fig:spe4ehflux} are close to 50\%.  Possible differences between tracking methods' results should not obscure our key point, though, which is that significant changes {\em occur}; while different methods might yield different magnitudes, we still expect they would {\em detect} such changes.  Replication of our analyses using other methods or other tracking parameter choices would be valuable, to assess the sensitivity of our results to these factors.

During major flares, the spectral line profile observed by HMI can be greatly deformed. In magnetogram pixels where this occurs, data are unreliable \citep{sun2017highres}.  This effect could impact our calculation of the helicity and energy fluxes near flare times. We evaluate its possible impact by using the 135-second-cadence vector data for AR 11158 with the bad pixels corrected by applying a step-like function to fit the time-series good data for each pixel before and after the flares, suggested by \citet{sun2017highres}. The result, shown in Figure \ref{hflux135s}, is  consistent with the aforementioned conclusion that the helicity flux decreases suddenly during the flares, followed by a recovery. This test suggests that the observed decrease in helicity and energy fluxes from the flares is real.  Further, the lack of any obvious dip in fluxes for confined events, which would be equally susceptible to any spectral-line artifacts, also implies this result is not spurious.

\begin{figure}[ht]
\centering
\includegraphics[width=0.50\textwidth]{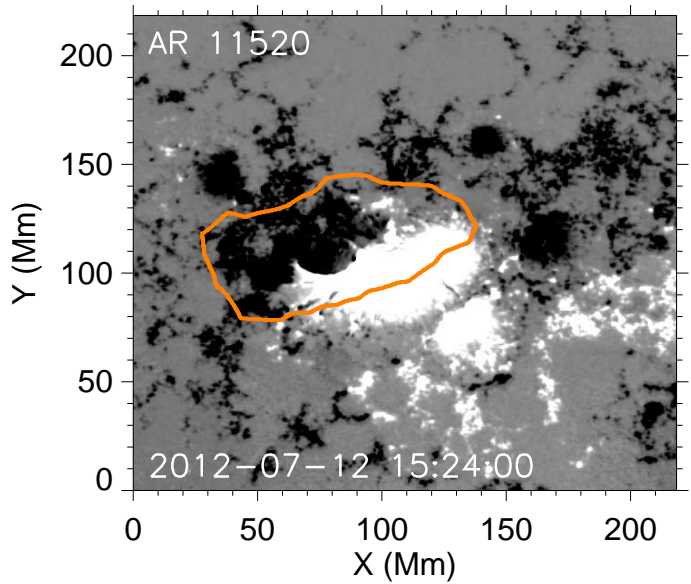}
\vspace{0.5cm}

\caption
{An example flare mask, for the X1.4 flare in AR 11520, is shown by the orange contour plotted over a magnetogram of $B_z$ (background grayscale) taken at 15:24 UT July 12, 2012. The flare mask's boundary is determined by combining the region's NLFFF-derived $Q$-map and flare ribbon/arcade emission.
}
\label{fig:flaremask}
\end{figure}

The decreases in helicity and energy fluxes after eruptive X-class flares must be related to changes in the magnetic and velocity fields upon which these fluxes depend. Indeed, it has been shown that the horizontal field in flare areas increases substantially after major flares (see, e.g., \citealt{Wang2010,sun2012,Petrie2019} for observational analysis, and, e.g., \citealt{Barczynski19} for modeling).
Changes in velocities associated with large flares have also been reported (e.g., \citealt{Deng2006}). To understand how these changes impact helicity and energy fluxes, we analyzed the distributions of magnetic fields, velocity fields, and energy flux densities from areas that participated in eruptive flares.  While the flux of helicity is gauge-invariant, the helicity flux {\em density} is not \citep[e.g.,][]{Pariat2005}, so we do not investigate changes in this quantity.

As a first step, we produced a so-called flare mask for each event, to isolate areas associated with the flare. Eruptive ARs, especially complex ones, often contain substantial flux that does not erupt, and connects to other polarities within the AR, to neighboring ARs, or to remote areas on the Sun. Because random changes from such irrelevant flux systems could conceal the variations associated with flares that interest us, we restrict our study of changes in ${\bf B}, {\bf V},$ and energy flux densities (Poynting fluxes) as much as possible to the areas from which fields erupt. To do so, we selected areas for each flare mask by visual inspection of both maps of the squashing factor $Q$ \citep{tit07,tit11} and the observed locations of flare arcades and ribbons. We first calculate $Q$ at the lower boundary using the NLFFF model. We then choose the high-$Q$ contours that enclose the flare patterns of arcades and ribbons. Figure \ref{fig:flaremask} shows a flare mask for the X1.4 flare in AR 11520 at 15:37 UT, July 12, 2012. The flare mask, denoted by the orange contour, is plotted on an $B_z$ magnetogram taken at 15:24 UT. Details about procedures used to generate a flare mask, as well as the masks for well-known ARs 11158, 11429, and 12192, are discussed by \citet{liu2017}. In most cases, this provides a good proxy for the area containing erupting flux. In each flare mask, we set the selected areas to unity and remaining areas to zero.   We then applied each flare mask to maps of the pre- and post-flare magnetic fields, velocity fields, and energy flux densities to extract all flare-involved pixels for all the eruptive events.  Pre-event maps were the last pre-flare observation; post-event maps were the first observation at least 30 minutes after the flare end time in the GOES catalog. 

\begin{figure}[ht]
\centering
\includegraphics[width=0.80\textwidth]{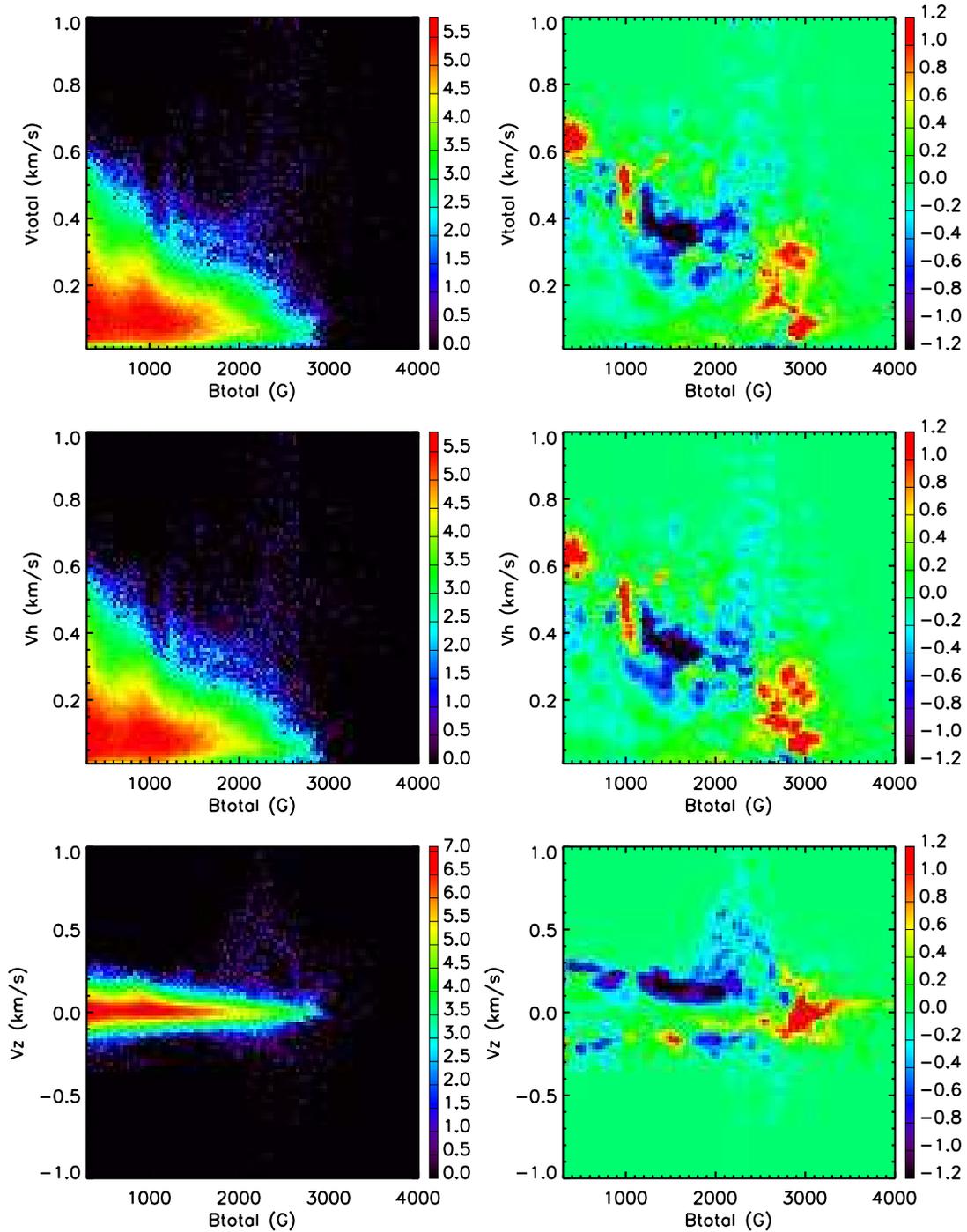}
\vspace{0.5cm}

\caption
{
Left column: Distributions of the logarithm of pixel counts in the 2-variable phase space of magnetic field versus pre-flare velocity field for all eruptive events. Right: Distribution of differences in pixel counts in the same 2-variable space, given by $\log$(post-flare counts) - $\log$(pre-flare counts).
Only pixels within the flare masks are included in these distributions.}
\label{fig:densitydifferencemap}
\end{figure}

In the left column of Figure  \ref{fig:densitydifferencemap}, we show the pre-flare distributions of the logarithm of pixel counts in the 2-variable phase space of magnetic field strength and 
total velocity (top), horizontal velocity (middle), and vertical velocity (bottom), before the flares. In the right column, we show distributions of {\em differences} in pixel counts, i.e., $\log$(post-flare counts) - $\log$(pre-flare counts), in each 2-variable space. In the right column, red indicates that more pixels occupy a given location after the flare in this 2-variable space, blue indicates that fewer do (and hence that more pixels occupied this given location before the flare).  The color bars saturate at logarithmic values of -1.2 to +1.2 for the post-to-pre ratio, corresponding to changes in occupancy by factors of exp(-1.2) $\simeq 0.30$ and exp(+1.2) $\simeq 3.3$.  In these distributions of changes associated with flares, we see clear increases in strong-field pixels with relatively low $|{\bf V}_h|$ and
$|V_z|$ (red features near 3000 G) 
and decreases in moderate-strength pixels with moderate speeds (blue features near 2000 G). 
An increase in weak-field pixels with larger $|{\bf V}_h|$ can also be seen.  Because stronger fields suppress convection more effectively than weaker ones (e.g., \citealt{Berger1998,Welsch2012}), it is plausible that the increases in horizontal field strengths that often occur in strong flares shifted pixels from the central blue patches to the strong-field red patches.

\begin{figure}[ht]
\centering

   \centerline{\hspace*{0.010\textwidth}
               \includegraphics[width=0.45\textwidth,clip=]{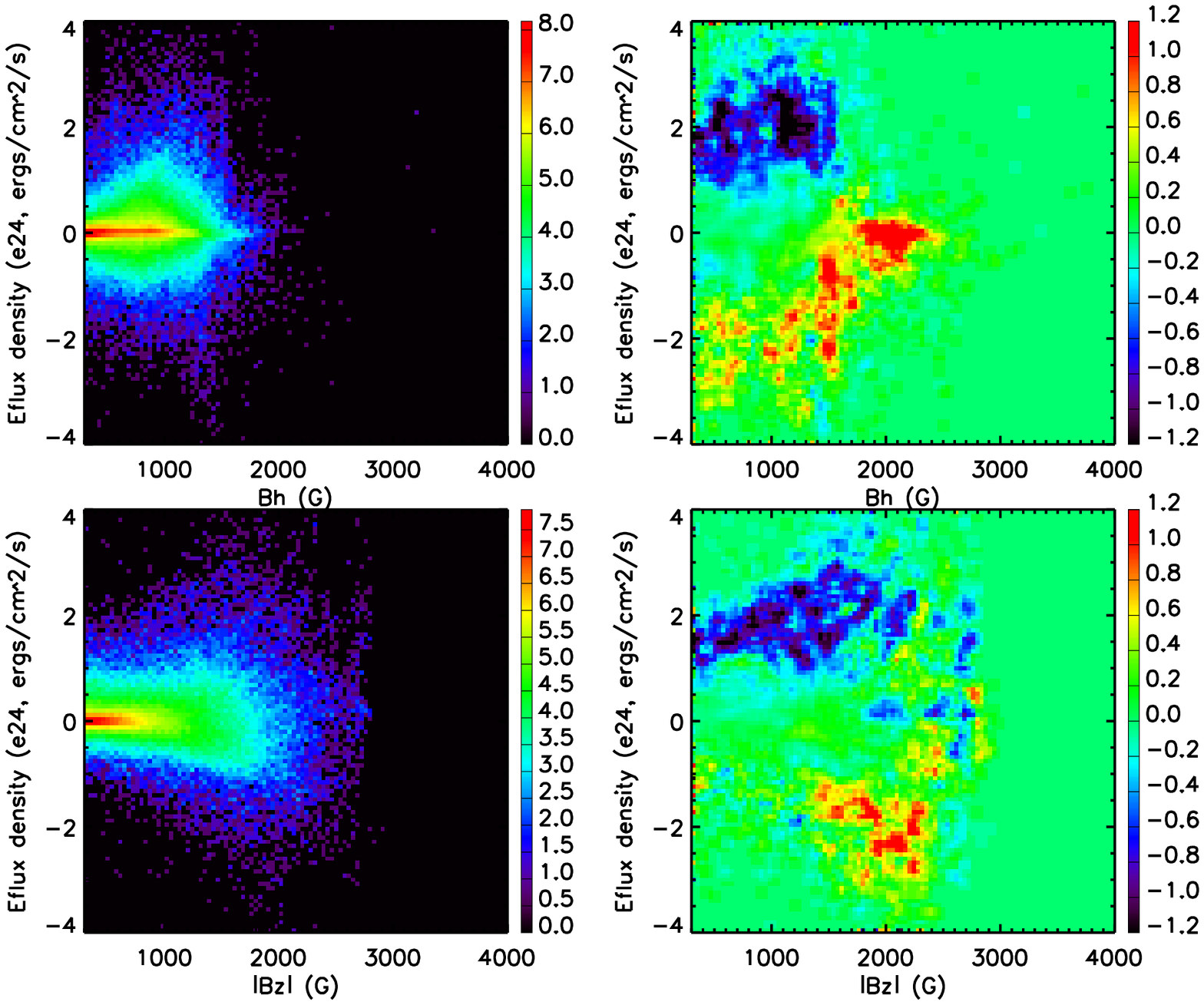}
               \hspace*{-0.0005\textwidth}
               \includegraphics[width=0.45\textwidth,clip=]{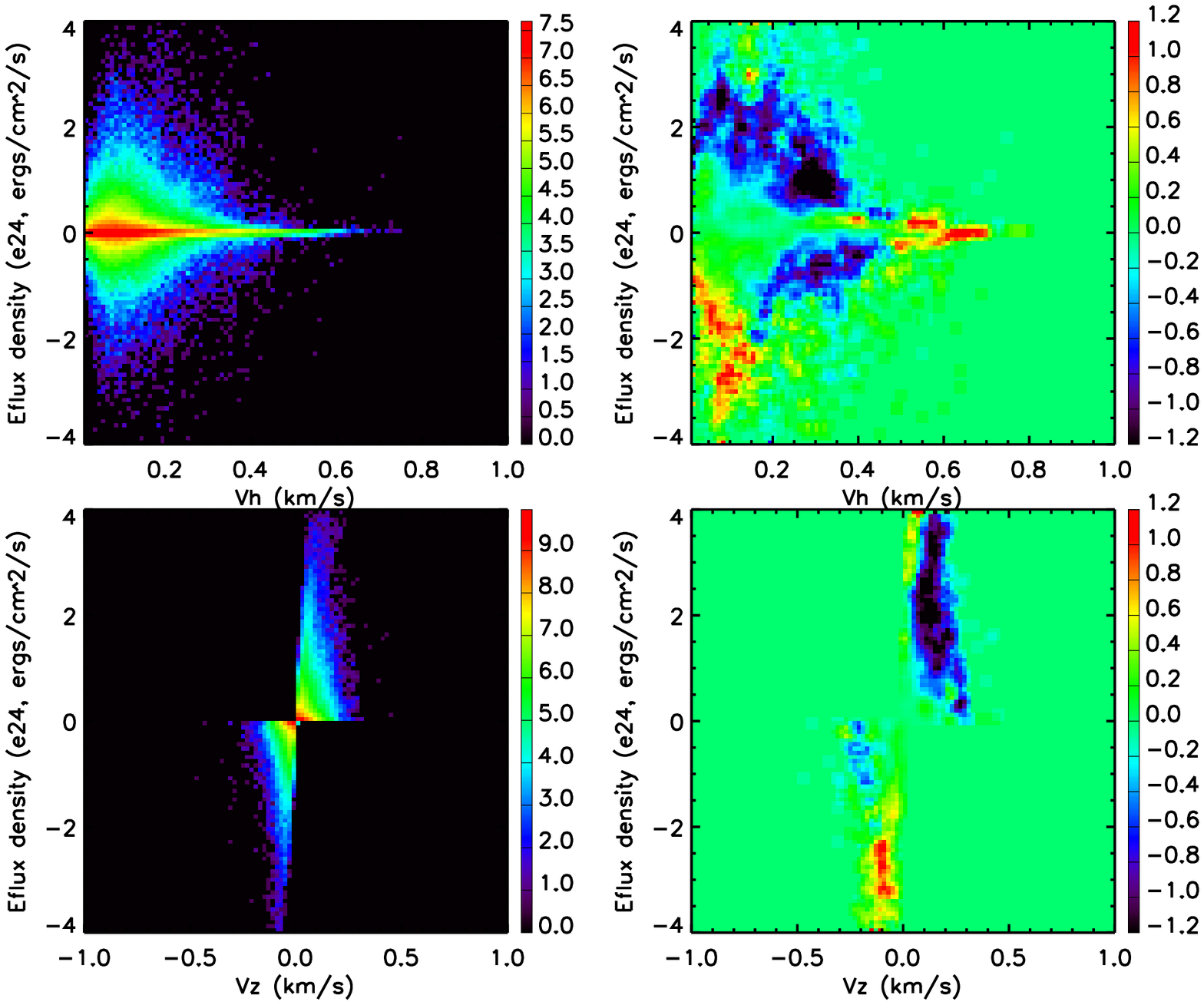}
              }
        
\caption{Left two columns: Distributions of the logarithm of pixel counts in the 2-variable phase space of magnetic field versus Poynting flux for all eruptive events (first column), and distribution of differences in pixel counts in the same 2-variable space, given by $\log$(post-flare counts) - $\log$(pre-flare counts) (second column). The top row shows the logarithm of counts and the logarithm of change in counts for $|{\bf B}_h|$, while the bottom row shows these quantities for $|B_z|.$
Right two columns: Distributions of the logarithm of pixel counts in the 2-variable phase space of velocity field versus Poynting flux for all eruptive events (third column), and distribution of differences in pixel counts in the same 2-variable space (fourth column).  The top row shows the logarithm of counts and the logarithm of change in counts for $|{\bf V}_h|$, while the bottom row shows these quantities for $V_z.$ Only pixels within the flare masks are included in these distributions.}
\label{fig:densitydiff4energy}
\vspace{-2mm}
\end{figure}

As shown in Figure \ref{fig:densitydiff4energy}, we also analyzed pre- and post-flare distributions of pixel counts in 2-variable phase spaces of Poynting flux versus 
$|{\bf B}_h|$ (top row) and $|B_z|$ (bottom row) in the first and second columns, and Poynting flux versus $|{\bf V}_h|$ (top row) and $V_z$ (bottom row) in the third and fourth columns. For each 2-variable space, the first and third columns show the logarithm of pre-flare counts, and the second and fourth columns show the logarithm of the difference in counts, post-flare minus pre-flare.  Panels in the second column show decreases in upward Poynting flux from moderate-strength pixels (blue areas around 1000 G) and increases in downward Poynting flux in higher-strength pixels (red areas around 2000 G). 
We note three features in the top panel of the fourth column: decreases in both positive and negative Poynting flux from pixels with moderate horizontal flows (blue areas around 0.3 km s$^{-1}$); a decrease in positive Poynting flux from pixels with weak horizontal flows (blue area near $|{\bf V}_h| \simeq$ 0); and an increase in negative Poynting flux from pixels with similarly weak horizontal flows (red area near $|{\bf V}_h| \simeq$ 0).   The bottom panel of the fourth column shows a large decrease in positive Poynting flux from pixels with upward flows.

The dominant shift of the pixel distributions in Figure \ref{fig:densitydifferencemap} was an approximate doubling of some horizontal field strengths and an approximate halving in flow speeds for that shifted population. In principle, these shifts might have offset each other, leading to little change in the energy flux. Figure \ref{fig:densitydiff4energy} demonstrates, however, that there is a change in the {\em character} of the Poynting flux, not just its magnitude.  First, more post-flare energy flux is downward. Second, the field strengths for which most changes in the Poynting flux occur (around 1000 -- 2000 G for both $B_h$ and $|B_z|$) differ from the field strengths at which most changes in flows occurred in Figure \ref{fig:densitydifferencemap} (around 1500 -- 3000 G for both $B_h$ and $|B_z|$).  
Because Poynting fluxes depend upon flow direction in addition to speed, these changes indicate that both directions and magnitudes of flows change after flares. 
Our analysis thus far, however, has not yet revealed the physical basis for such changes.  
In a separate study, currently underway, we are further analyzing the nature of these changes in flows and fields.

As remarked earlier, the helicity flux density is a gauge-dependent quantity, and thus does not have a clear physical meaning. It is therefore uncertain what causes the decrease of helicity flux after flares. The decrease in helicity flux seen in Figure \ref{fig:spe4ehflux} accords with the decrease in strong fields' horizontal speeds  after flares shown in the right column of Figure \ref{fig:densitydifferencemap}. Like the Poynting flux, however, the helicity flux depends on flows' directions in addition to their magnitudes, so we cannot rule out the possibility that the same changes in flows' directions that increased downward energy transport also reduced the net injection of helicity into the corona. 

\subsection{Flare-Induced Changes in Coronal Helicity and Energy}
\label{ss:changeHE}

\begin{figure}[ht]
\centering

   \centerline{\hspace*{0.015\textwidth}
               \includegraphics[width=0.45\textwidth,clip=]{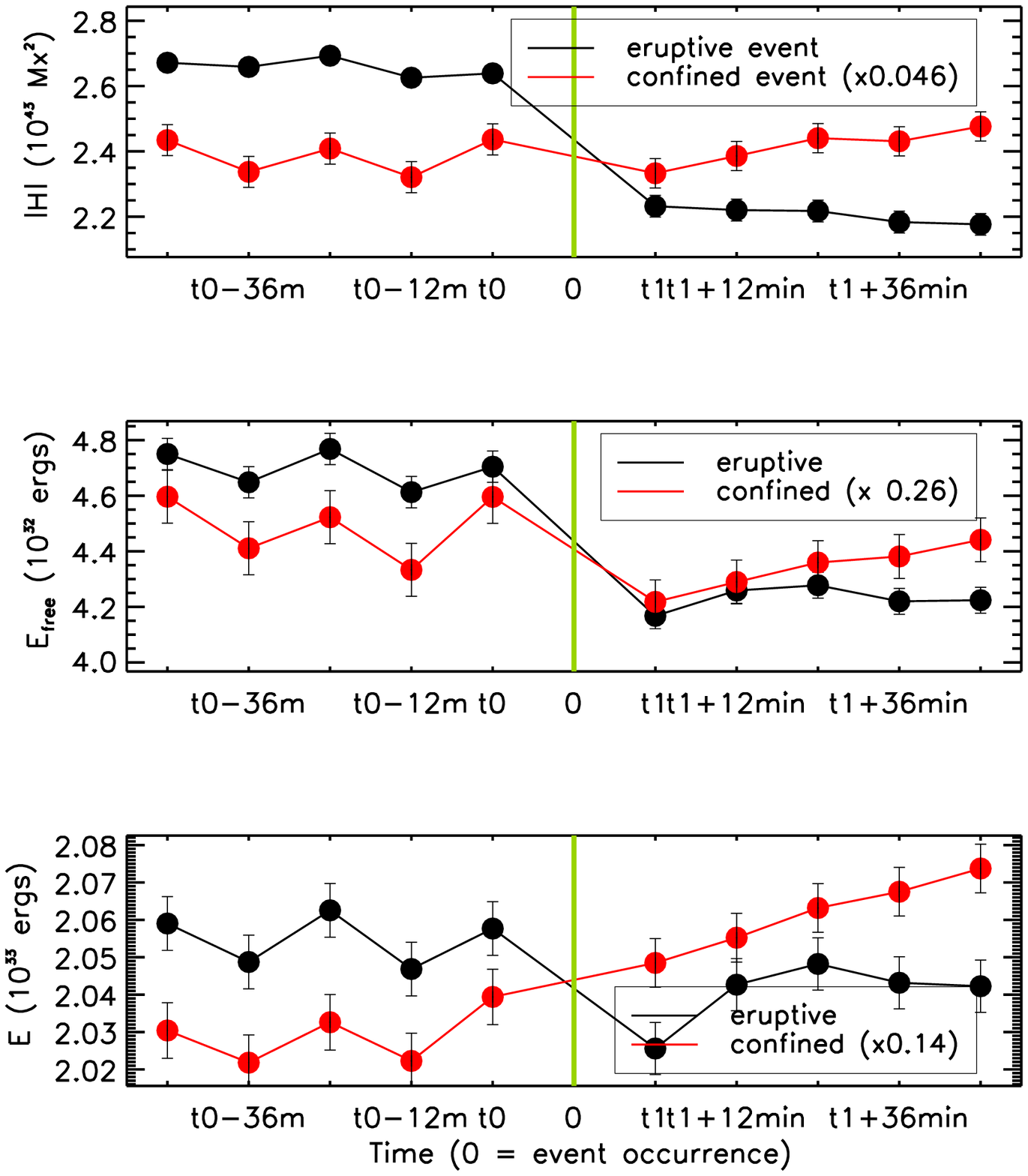}
               \hspace*{-0.005\textwidth}
               \includegraphics[width=0.45\textwidth,clip=]{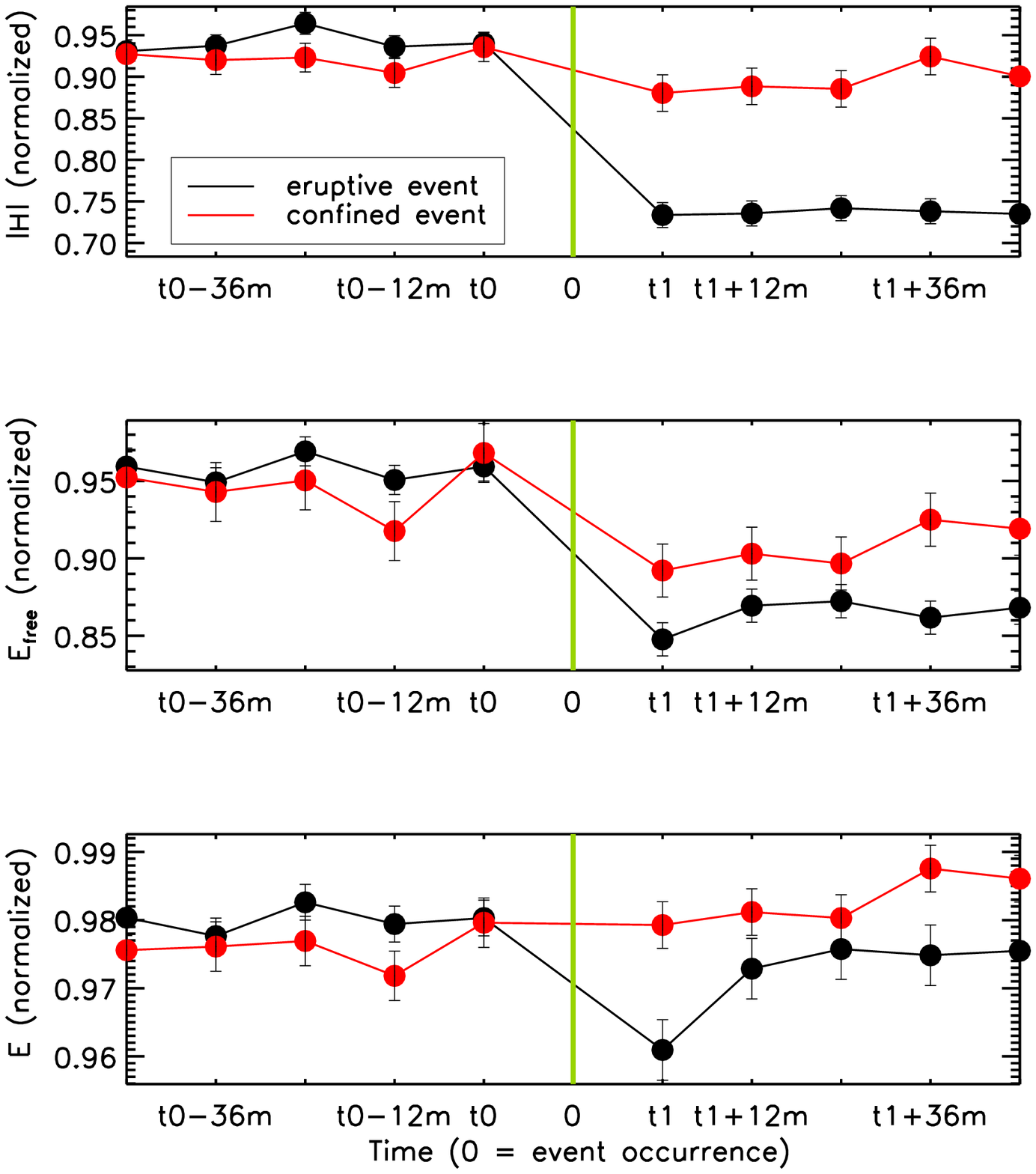}
              }
\caption{Superposed epoch plots of unsigned magnetic helicity (top), free magnetic energy (middle), and total magnetic energy (bottom) in the active regions for the eruptive (black) and confined (red) X-class flares. The vertical green lines denote the flare start time. Five data points immediately before the flares and five data points starting 30 minutes after the end of the flares are chosen for these calculations. The time before the flares is denoted as t0; the time 30 minutes after the end of the flares is denoted as t1. Left column: Time series in physical units combined via SPE. Right column: Normalized time series combined via SPE. Note that the horizontal axis shows step number (not time) before / after each event; the time step in each set of five points is 12 minutes.}
\label{fig:speHandE}
\vspace{-2mm}
\end{figure}

Having analyzed changes in helicity and energy {\em fluxes} associated with our sample of X-class flares, we next study evolution in the {\em coronal content} of helicity and energy associated with these flares.
As shown in Figure \ref{fig:example11158}, the helicity present in our model of the coronal field of AR 11158 decreases after the eruptive X2.2 flare.  To determine if coronal models of other ARs in our sample tend to exhibit similar flare-associated changes, we created SPE plots, shown in Figure \ref{fig:speHandE}, of magnetic helicity (top), free energy (middle) and total energy (bottom) for the eruptive (black) and confined (red) events. 
The left-column SPE plots were made combining events' time series in physical units.   Because these quantities for each AR scale with that AR's total unsigned flux, evolution in a subset of high-flux ARs could dominate these curves.   To check this, we also created the right-column SPE plots by first normalizing each AR's time series by its highest value over the entire interval that it was observed and then combining the regions' time series.

In each plot, we show five data points before and five data points after the event times, i.e., the GOES flare start and end times in Table \ref{table:xflares}. The data cadence for each five-point block is 720 seconds. The $t0$ in the x-axis refers to the flare start time. To avoid bad data immediately after these major flares, the post-event data start with the first HMI observation at least 30 minutes after the end time of each flare. The $t1$ in the x-axis refers to this time. The green vertical line denotes the flare time, but its placement with respect to pre- and post-flare data points should not be viewed as physical: due to differences in flares' start times and durations with respect to HMI's nominal, 720-second cadence for vector magnetograms, the spacing between the last pre-flare and first post-flare data points shown does not correspond to 1440 seconds.  
Our use of variable-sized time windows is appropriate, given the physical basis for our study: we believe that magnetic reconnection is the ultimate cause of both (i) the GOES emission that determines the events' start and stop times and (ii) the changes in magnetic energy and helicity (and their fluxes) that we analyze.  

\begin{figure}[ht]
\centering
\includegraphics[width=0.50\textwidth]{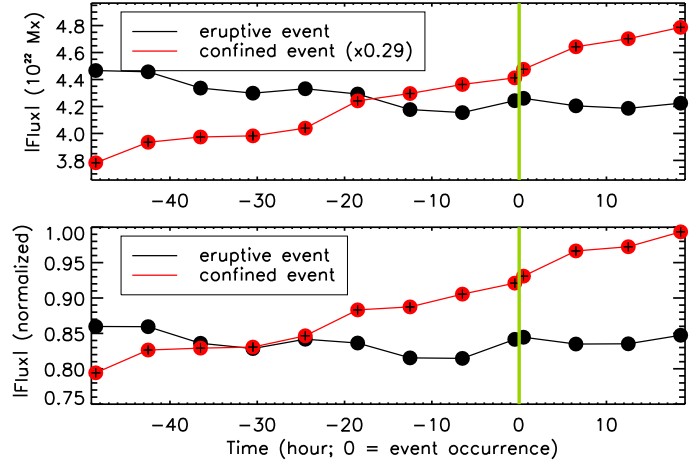}
\vspace{0.5cm}

\caption
{Superposed epoch plots of total unsigned (TUS) magnetic flux (top) and normalized magnetic flux (bottom) in the active regions for eruptive (black) and confined (red) X-class flares. The vertical green lines denote the flare occurrence time. The absolute flux in all pixels with field strength greater than 300 G was summed for these magnetic flux estimates, i.e., flux in both polarities was included.
}
\label{fig:SPE4Mflux}
\end{figure}

Flare-associated changes are noticeable for eruptive flares in all six plots, and for confined events in the two versions of the free energy plot.  To assess the significance of these changes, a uniform uncertainty was computed for the pre-flare epoch, and, separately, for the post-flare epoch, estimated from the standard deviations of the corresponding five data points. 
The error bars shown in the SPE plots are calculated from the propagated errors, i.e., as the square root of the mean of the total variance (the sum of squared uncertainties in the quantities being averaged).  The change in each quantity (in physical units) in the left-column plots was computed as the average of the five post-flare points minus the average of the five pre-flare points.  For eruptive events, both magnetic helicity and free energy decrease by significant amounts after the flares. In contrast, only free energy is seen to decrease after the flares for the confined events, and the magnitude of its decrease from the pre-flare average is only significant at about the 1-$\sigma$ level. For the eruptive events, helicity decreases by about $(0.45\pm 0.04)\times 10^{43}$ Mx$^2$ (about 17\% of the pre-event average), and free energy decreases by about $(0.47 \pm 0.07)\times 10^{32}$ ergs (about 10\% of the pre-flare average).  The decrease in total magnetic energy for eruptive events is only marginally statistically significant at the 1-$\sigma$ level, and at roughly  $(0.15\pm 0.10)\times 10^{32}$ ergs, it is only about 0.7\% of the pre-flare average. The fractional decrease in helicity is about twice as large as that in free energy, implying that helicity is more sensitive to changes in eruptive systems than free energy.  For confined events, the decrease in free energy is about $(0.59 \pm 0.48) \times 10^{32}$ erg.  In confined events, the other quantities actually increase after flares, by 1.1\% for total helicity and 1.6\% for total energy (Figure \ref{fig:SPE4Mflux}, discussed below in our analysis of the ARs' longer-term evolution, is also relevant here.) 
However, this 1.1\% increase in total helicity may not be significant because computational errors can be larger than this change.

\begin{figure}[ht]
\centering
\includegraphics[width=0.70\textwidth]{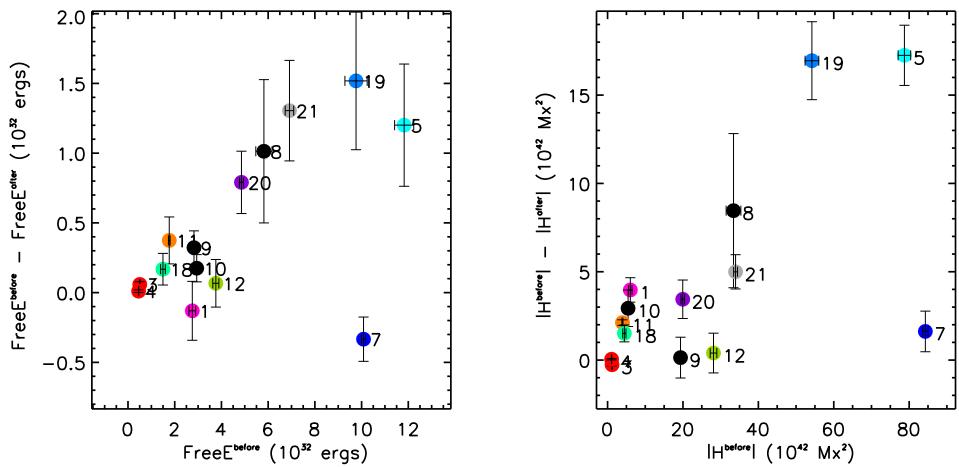}
\vspace{0.5cm}

\caption
{
Changes in free energy (left) and magnetic helicity (right) in eruptive events, plotted against ARs' available free energy and helicity, respectively. Each flare is denoted by its flare ID listed in Table \ref{table:xflares}. Because flares 5 and 6 were less than an hour apart, the two are treated as one
event here, plotted at the point labeled 5.}

\label{fig:changeHE4eruptiveflares}
\end{figure}

We also find evidence that both the free energy and helicity {\em released} in eruptive flares tend to be larger when more of these quantities were {\em available} to be released, as shown in Figure \ref{fig:changeHE4eruptiveflares}. 
This is analytically suggested in linear theory \citep{valori2015}.
Figure \ref{fig:changeHE4eruptiveflares} may also suggest upper limits of $\sim$20\% of the free energy and $\sim$30\% of the total helicity in the ARs to be released by major eruptive flares. This is also consistent with estimates from previous studies using both observational and synthetic data \citep[see, e.g.,][]{Nindos2003,Moraitis2014}. 

The difference in free energy, $E_{\rm free, before} - E_{\rm free, after}$, is expected to be positive if energy is released.  The negative change estimated for one event could arise from uncertainties in the NLFFF models, or from other evolution in parallel, such as flux emergence, during the pre- to post-flare interval. This relationship need not imply causality in the reverse direction, with {\em availability} of more energy and helicity {\em driving} the release of more, as opposed to simply {\em enabling} the release of more.  Based on observations of a nearly ``universal'' power-law distribution of flare number frequency versus energy for solar ARs (e.g., \citealt{Crosby1993,Aschwanden2016} - but see also \citealt{Wheatland2010} for a possible exception), a reasonable prior expectation is that the free energy released in a given event is a random variable drawn from a power-law parent distribution. The low free-energy-loss and low helicity-loss outlier in the lower right corner of each panel is consistent with this expectation of stochasticity, subject also to the specific local conditions where the flare was triggered.  
An alternative model is the build-up-release (BUR) scenario, described by \citet{Hudson2020}, which predicts that available free energy in ARs is released as it is accumulated.  The BUR view appears inconsistent with our observations in that released free energy is a small fraction of available free energy in all cases.    
Whatever the underlying physical mechanisms governing free energy and helicity release, the correlations here could have implications for space weather prediction: when more energy is available, more tends to be released, with more repercussions for associated or secondary phenomena, such as CMEs and solar energetic particle events.

\begin{figure}[ht]
\centering

   \centerline{\hspace*{0.005\textwidth}
               \includegraphics[width=0.44\textwidth,clip=]{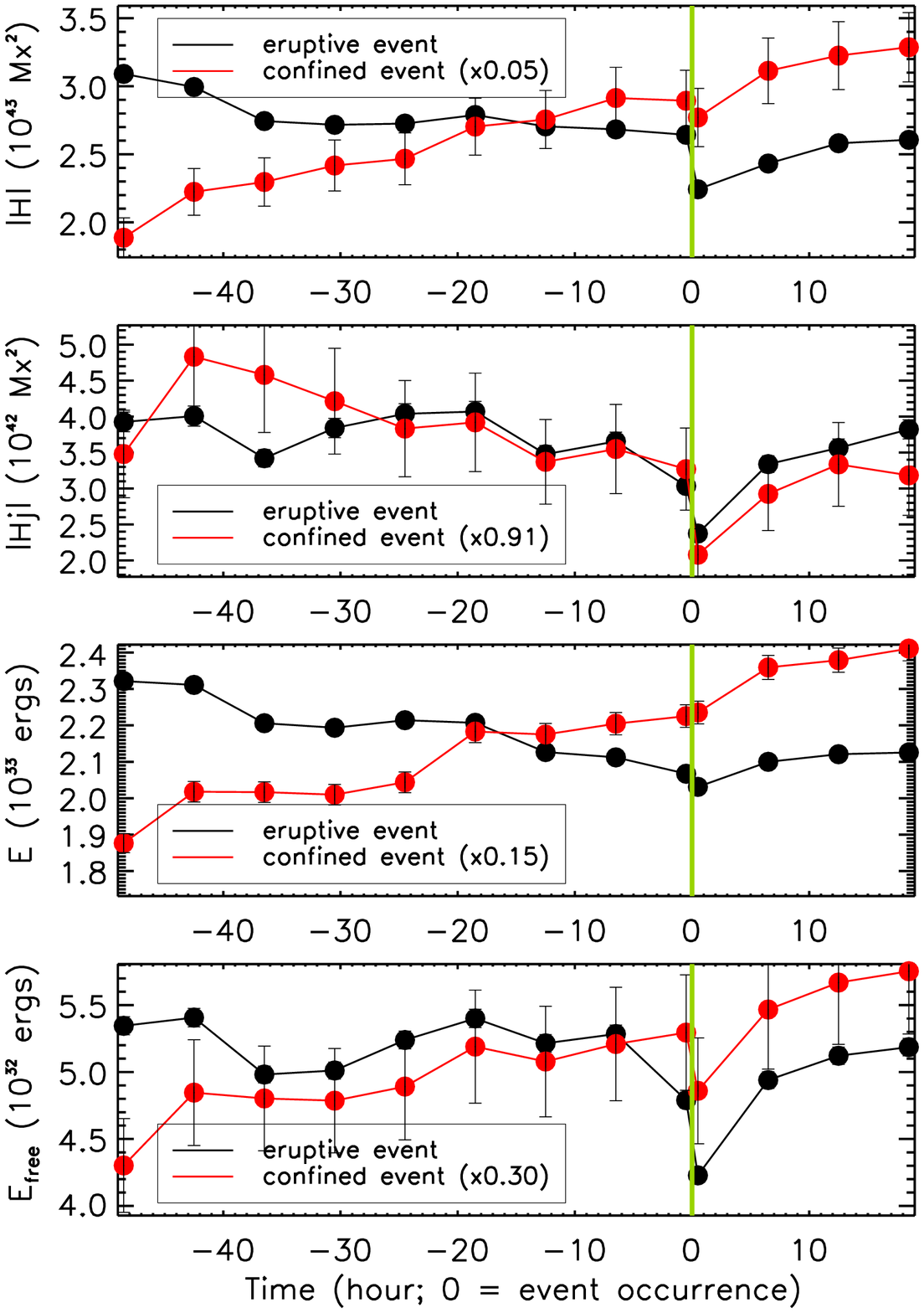}
               \hspace*{-0.005\textwidth}
               \includegraphics[width=0.45\textwidth,clip=]{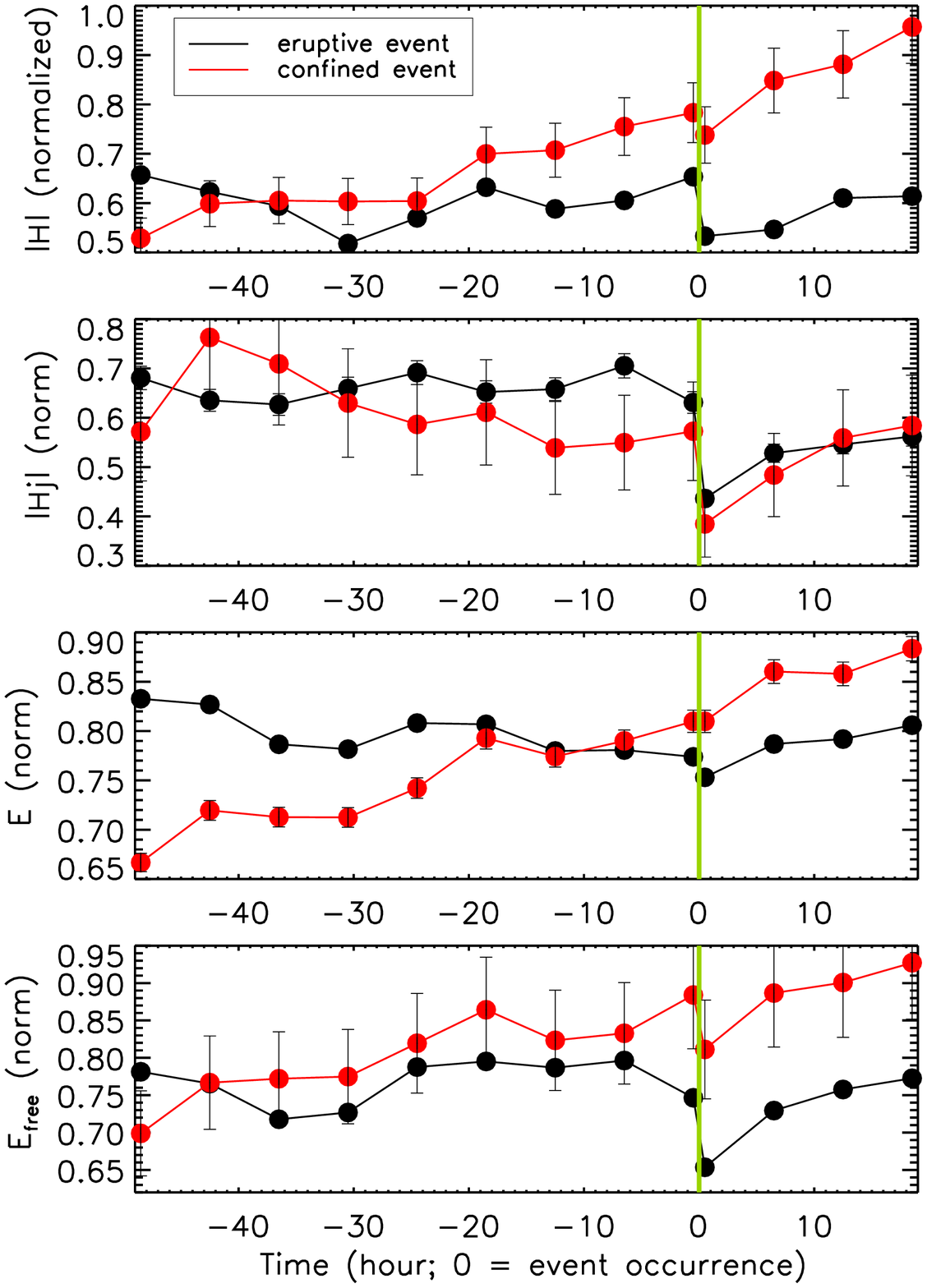}
              }
\caption{Same as Figure \ref{fig:speHandE} but for longer time intervals. The temporal profile of helicity of the current-carrying field ($|H_j|$), one component of the helicity, is also plotted in the second row. 
Left column: Time series in physical units, combined via SPE. Right column: Normalized time series, combined via SPE.}
\label{fig:SPE4H_evo}
\vspace{-2mm}
\end{figure}

The SPE plots in Figures \ref{fig:spe4ehflux} and \ref{fig:speHandE} show aspects of magnetic evolution close to the times of X-class flares. What can we learn about such evolution from SPE plots over longer time scales before and after these flares?   Figure \ref{fig:SPE4H_evo} shows SPE plots from 48 hours pre-flare to 18 hours post-flare, for unsigned helicity, $|H|$ (first row), unsigned helicity in the current-carrying field, $|H_j|$ (second row), total energy, $E$ (third row), and free energy, $E_{\rm free}$ (fourth row). As above, we combined ARs' time series in both physical units (left column) and after normalizing each AR's time series (right column) by that AR's peak value over the interval of its observation.

In these longer duration SPE plots, it should be noted that, due to our small sample of confined events, some observations are included more than once: for the same AR, an observation $\Delta t_1$ hours before one flare can also be $\Delta t_2$ hours before a second flare, and the synchronization of flare times can include both when a long enough pre-flare interval is plotted. This overlapping of multiple segments of an AR's time series can, by summing evolution at different phases, possibly obscure significant evolution that the SPE approach is meant to highlight.  Effects from shifting and combining multiple observations from a single AR could even produce trends that are spurious.  Consequently, one expects stronger conclusions to be achieved by applying SPE analysis to a larger sample of confined flares with no duplicate epochs used in the superposition.   

With this caveat in mind, few pre-flare trends in these quantities are evident.  Prior to confined events, most of these quantities increase monotonically, but the small number of confined events in our sample must be kept in mind.  
For eruptive events, no buildup of energy and helicity prior to these X-class flares is obvious, but there is a clear tendency in all four quantities for post-flare evolution {\em toward} pre-flare levels over the first 12 hours after each event.  In particular, the helicity of the current-carrying field and free energy appear to be essentially replenished on this time scale. For smaller flares the replenishment might happen on shorter timescales, see e.g., \cite{2021A&A...652A.159D,2022A&A...T}.

Because the emergence of new flux can affect all of the quantities plotted in Figure \ref{fig:SPE4H_evo}, and might trigger flares/CMEs (e.g., \citealt{Priest1974}), it is worthwhile to also investigate the evolution of total unsigned (TUS) flux in our AR ensemble. Flux emergence should cause increases in TUS flux, and flux cancellation (another possible trigger of CMEs; e.g., \citealt{Amari2010}) should cause decreases in TUS flux. 
We remark that flux cancellation might not be discernible in ARs in which flux emergence is also occurring; but superposing evolution of TUS flux in our sample regions might reveal it.
For each AR, we computed TUS flux in the [-48,+18]-hour interval around each flare by summing the absolute value of radial field in pixels with total field strength greater than 300 G (note that this is approximately twice the flux in a single polarity.)   We then superposed the TUS flux curves for all epochs.  Our results are shown in Figure \ref{fig:SPE4Mflux}.  The top panel shows summed epochs for time series in physical units; the bottom panel shows summed epochs for normalized data. Error bars are smaller than the plotted circular symbols.
For eruptive events, no strong trends are visible, though a slight  increase in flux tends to occur during the 8 hr prior to event onset.  
The physical significance of this minor increase is unclear. For confined events, a monotonic increase in TUS flux is seen over the entire plotted interval. Hence, ongoing flux emergence is a plausible explanation for the increases in total helicity, free energy, and total energy for confined flares seen in Figure \ref{fig:SPE4H_evo}, as well as the slight increases of total helicity and total energy near flare times seen in Figure \ref{fig:speHandE}.

\subsection{Flare-Induced Changes in Magnetic Twist}
\label{ss:twist change}

Non-potentiality of magnetic fields in ARs can be quantified by various indices \citep[e.g.,][]{Leka2003a}. Here, we use magnetic twist as a measure of non-potentiality to investigate magnetic changes in ARs associated with X-class flares.  Based on the force-free field assumption, 
\begin{equation}
    \bf{\nabla} \times \bf{B} = \alpha \bf{B} ~,
\end{equation} 
the force-free parameter, $\alpha$, is often treated as a measure of magnetic twist in an AR \citep[e.g.,][]{seehafer1990,pevtsov1995}.
A scalar, $\alpha$ varies in space, and is often estimated for each pixel from vector magnetograms via 
\begin{equation}
    \alpha = [{\bf \nabla \times B} ]_z/B_z ~, 
     \label{eqn:alpha_by_pix}
\end{equation} 
using finite-difference approximations. To characterize twist in an entire AR, we choose a $B_z^2$-weighted, force-free 
$\alpha _w$, proposed by \citet{hagino2004}, 
\begin{equation}
\alpha _w = \frac{\displaystyle \int_{S} \alpha(x, y) \, B_z^2(x, y) \, dx \,dy}{\displaystyle \int_{S} \, B_z^2(x, y)\, dx \, dy},
\label{eqn:alphaav} 
\end{equation}
where $B_z$ is the vertical component of magnetic field. The integral is done over the region of interest in the AR. One advantage of $\alpha_w$ over a simple average of force-free $\alpha$ from Equation (\ref{eqn:alpha_by_pix}) is that singularities at the polarity inversion lines (where $|{\bf B}|$ can be large but $B_z$ is zero) are avoided \citep{tiwari2009}. Because the $B_z$ weighting de-emphasizes values of $\alpha$ from noisy, weak-field pixels,  $\alpha_w$ is also believed to be one of the most robust approaches in computing $\alpha$ \citep{leka1999}.   We note that $\alpha_w$ is an {\em intensive} parameter -- that is, it does not increase with an AR's flux content or size.   
This is in contrast to $E, dE/dt, H,$ and $dH/dt$, which are {\em extensive} parameters, i.e., they tend to increase with system size.  We remark that a high-flux active region need not, in principle, also be highly non-potential. In practice, however, \citet{Fisher1998} found that, in a sample of a few hundred ARs, totals of unsigned vertical current in ARs were highly correlated with their magnetic fluxes.  
Based on this observation, we expect integral measures of non-potentiality to scale with AR size.  We also note that some parameters, or combinations of parameters, might not be readily classified as either extensive or intensive.
\citet{Leka2007, Welsch2009, Tan2009}, and \citet{Bobra2015} found that extensive magnetic parameters tended to be better predictors of flare productivity than intensive parameters.  In contrast, \citet{Bobra2016} found that intensive parameters were better predictors of CME occurrence.  

\begin{figure}[ht]
\centering
\includegraphics[width=0.60\textwidth]{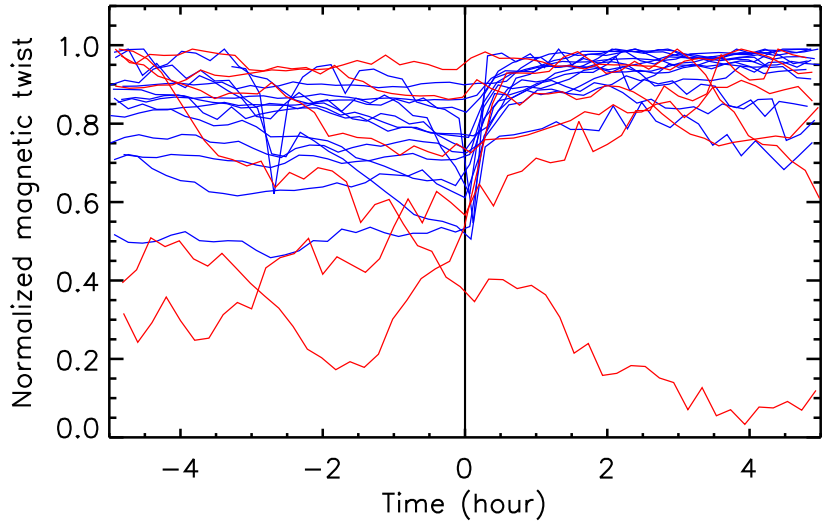}
\vspace{0.5cm}

\caption
{
Temporal profiles of normalized magnetic twist near the times of X-class flares in our AR sample. Blue curves refer to eruptive flares; red confined flares. The vertical black line denotes the occurrence time of the flares. 
}
\label{fig:twist4all}
\end{figure}

\begin{figure}[ht]
\centering

   \centerline{\hspace*{0.015\textwidth}
               \includegraphics[width=0.48\textwidth,clip=]{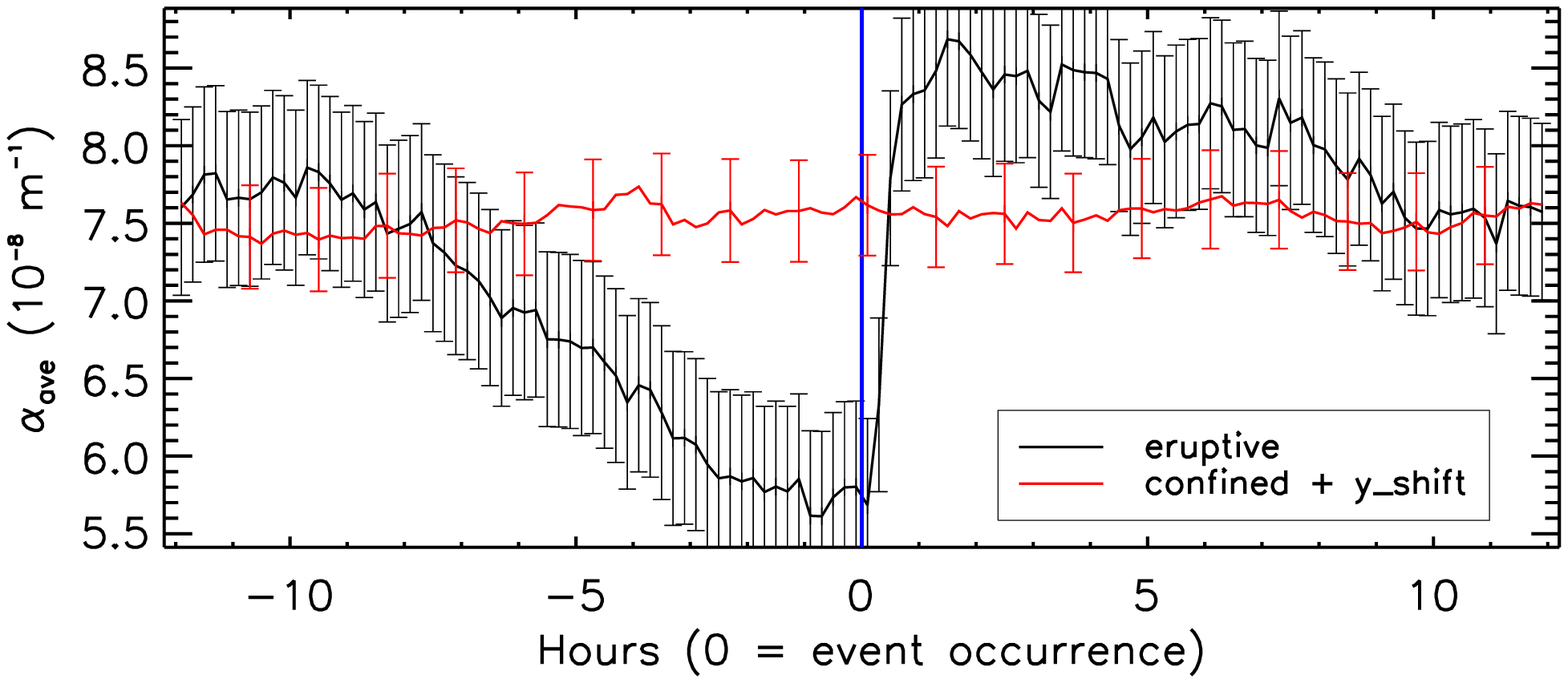}
               \hspace*{-0.0008\textwidth}
               \includegraphics[width=0.48\textwidth,clip=]{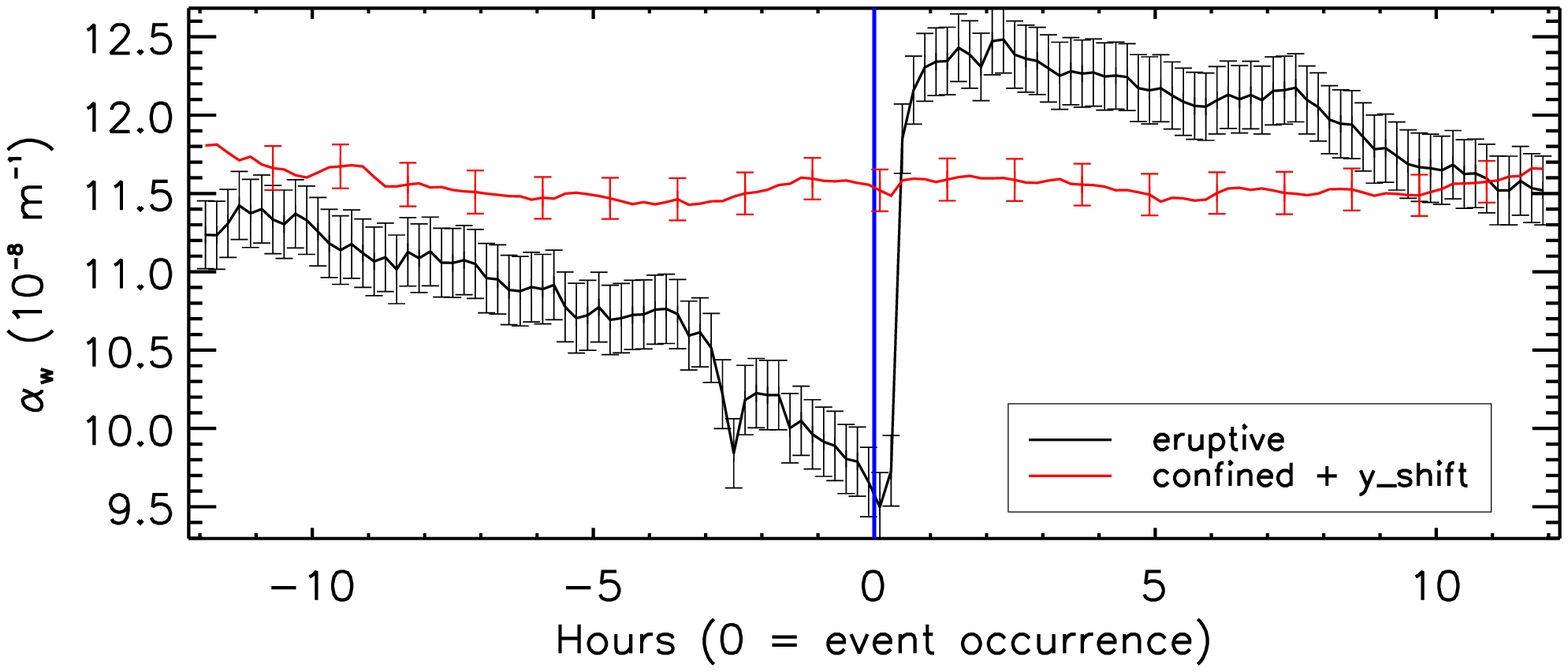}
              }
           
\caption{Superposed epoch plots of unsigned averaged magnetic twist from Equation (\ref{eqn:alpha_by_pix}) (left) and from Equation (\ref{eqn:alphaav}) (right). The pixels in the flare masks with field strength greater than 300 G are included for the calculations. The SPE for confined events is shifted upward along the y-axis by an amount y\_shift to aid comparison of the time evolution. y\_shift = $7.15\times 10^{-8}$ m$^{-1}$ for averaged $\alpha_{ave}$ (left) and $10.99\times 10^{-8}$ m$^{-1}$ for weighted averaged $\alpha_w$ (right). Blue vertical lines denote flare start time.}
\label{fig:spe4twist4twomethods}
\vspace{-2mm}
\end{figure}

Figure \ref{fig:twist4all} shows normalized temporal profiles of unsigned magnetic twist for the eruptive (blue) and confined (red) events. Sudden jumps occur at the times of the flares for most events. This sudden increase appears to be permanent: the twist stays at the increased level after the events. SPE plots of $|\alpha|$ for our event sample are shown in Figure \ref{fig:spe4twist4twomethods}, computed from either Equation (\ref{eqn:alpha_by_pix}) (left) or  Equation (\ref{eqn:alphaav}) (right). The flare masks are applied for both calculations, i.e., only the pixels in the flare masks are included for the computations. The same flare mask is applied for all time steps for each event.\footnote{
We have also tested the sensitivity of our results to variations in flare-mask boundaries by dilating the flare masks. The time series $\alpha_{ave}$ and $\alpha_w$ retain very similar shapes, though the magnitudes of their changes in eruptive flares decrease with the level of expansion of the mask, up to 46\% with a half-width of 11 pixels for the dilation. The magnitudes of both $\alpha_{ave}$ and $\alpha_w$ themselves (cf., their changes) also become much smaller.
We speculate that expanding the masks adds pixels that are unrelated to the high-twist fields near the flaring PILs, with the added $\alpha$ values randomly distributed in sign and magnitude. Thus, the added pixels only increase the denominators in these averages.
}
With either approach, eruptive (black) events show a significant twist increase after the occurrence of the flares. We remark that twist in confined events is about one order of magnitude smaller than in eruptive events -- a striking difference.  Note that helicity {\rm flux} in confined events in Figure \ref{fig:spe4ehflux} is about 10 times {\em larger} than the helicity flux in eruptive events, but here we find the twist {\em content} in confined events is much {\em smaller} for confined events.  The SPE helicity flux time series for the set of confined events is dominated by AR 12192, which is a very high-magnetic-flux AR, making its $|dH/dt|$ very large.  This juxtaposition highlights the difference between the scaling of extensive parameters, like $dH/dt$, and intensive parameters, like $\alpha$. This accords with findings by \citet{Bobra2016}, who reported that high values of intensive twist measures were associated with CME occurrence. The uncertainty in average twist is large relative to fluctuations in twist for the confined flares, due both to our small sample of such events and the relatively low value of twist in those ARs (lower signal-to-noise).  Consequently, changes in twist after confined flares (e.g., a small increase in the right panel) are not significant. 

The increase in ARs' magnetic twist after flares must be caused by changes in their photospheric magnetic fields, most notably the enhancement of horizontal fields. Indeed, near flaring polarity inversion lines in strong flares, horizontal field strengths are often observed to increase, while vertical field strengths are not observed to change significantly \citep[e.g.,][]{wang2012, sun2012}. Magnetic shear along the PIL also often increases \citep{wang2012}. 
The increases in average twist might reflect the increased signal-to-noise in computing $\alpha$ values in stronger horizontal fields: horizontal fields below our 300G noise-floor threshold are excluded from our calculation, but the increase of horizontal field strengths after flares implies more pixels are included in post-flare magnetograms.
We note that the fractional increase in twist is larger in the left panel of Figure \ref{fig:spe4twist4twomethods}.  The $B_z^2$-weighting in Equation (\ref{eqn:alphaav}), for the right panel, possibly acts to suppress the contributions from pixels near PILs, where $|\mathbf{B}|$ is large but $|B_z|$ is small.  Hence, the difference between the fractional changes in these plots suggests that the magnetic changes responsible for much of the increase in twist occur in near-PIL fields.

The long-term trend of decreasing twist prior to events is significant, but we do not have a definitive explanation for it.  
Decreasing pre-flare twist might plausibly be an indication of expanding coronal magnetic structures (e.g., flux ropes) as coronal energy builds up. 
If the structure about to erupt gradually ascends in a metastable evolutionary stage, this could conceivably make the underlying photospheric fields less horizontal, with little or no change in $B_z$.  This scenario accords with observations of slow rising motions seen before some eruptions (e.g., \citealt{Schuck2004, Sterling2005, Liu2012f}). This is, in essence, an evolutionary version of the standard flare model in which the plasmoid and the X-point beneath it gradually ascend a few hours before the eruption.  Immediately after the eruption, near-PIL horizontal fields would become stronger from the implosion beneath  coronal reconnection sites, and, overall, values of the predominant, post-flare $\alpha$ would have higher magnitude. 
Unfortunately, the twist measures that we study are only photospheric, so do not provide unambiguous information about coronal magnetic structure.
We note that the flares masks over which these measures of average twist were derived {\em post facto}, and we have not analyzed any flare-quiet epochs, so it is unclear whether these trends could have any predictive capability. This finding may warrant futher work. 

An alternative interpretation of post-flare twist increases might involve the Taylor relaxation hypothesis \citep{taylor1974}, which predicts that magnetic reconnection should tend to make $\alpha(x,y)$ more uniform across ARs.  Physically, the Taylor hypothesis is less applicable to coronal fields than to laboratory plasmas, because it presumes that reconnection can operate on all fields present, and, unlike a laboratory system, a given coronal magnetic system is not truly isolated.  As \citet{Antiochos2003} and \citet{Longcope2008} have noted, however, photospheric line-tying of coronal fields implies that not all coronal fields can reconnect in a given flare.   Instead, reconnection can only act to make twist more uniform within the domain that participates in a given flare.  Also unlike laboratory systems, the corona is an open volume: when a flare is eruptive, some helicity in the source region is ejected into the heliosphere. Nonetheless, in general, helicity remains in post-reconnection fields that remain closed: indeed, magnetic shear along PILs is often observed to increase after major flares (e.g., \citealt{Wang2010}). In fact, this finding spurred a heated debate in the community when it was first reported \citep{Wang1994}.

In a study of several dozen ARs, \cite{Nandy2003} reported evidence for this effect in a subset of ARs whose twist was characterized for several successive days.  Their analysis included all GOES flares ascribed to each region in their sample. It is plausible that  multiple episodes of reconnection occurred throughout the ARs that they studied, enabling twist to become more uniform across these ARs over time.  By following the variance in regions' twist over several days, they estimated a characteristic timescale of circa eight days over which regions' twist tended to become more uniform.

\begin{figure}[ht]
\centering
\includegraphics[width=0.50\textwidth]{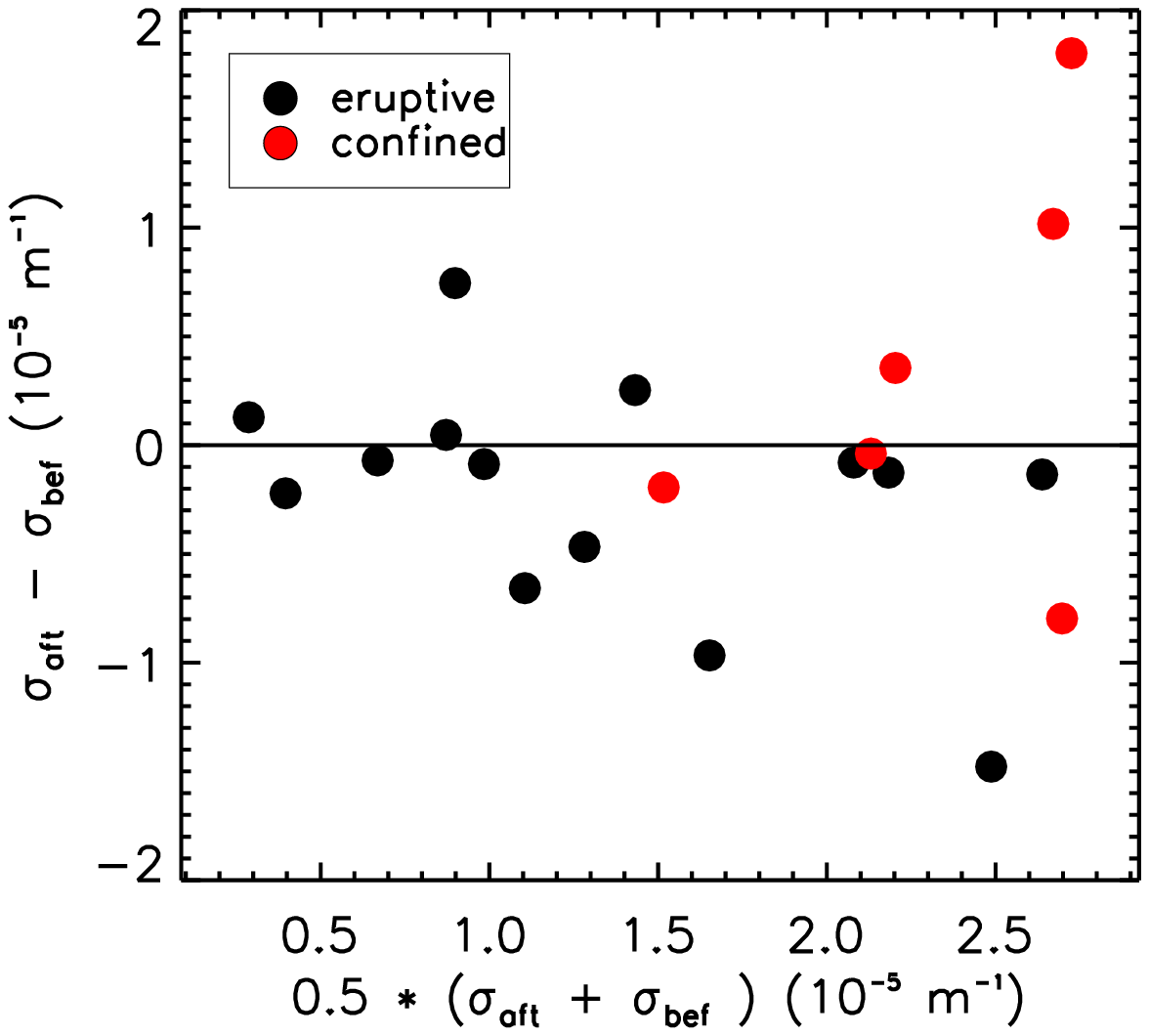}
\vspace{0.5cm}

\caption
{
Changes in the standard deviation of twist before and after 20 events, one difference per event, are shown for eruptive source regions (black) and confined source regions (red). Twists from pixels enclosed in the flare masks with field strengths greater than 300 G were included. 
}
\label{fig:twistVariancewithmask}
\end{figure}

To see whether twist within our AR sample became more spatially uniform after our flares, we computed the difference (post-flare minus pre-flare) in the standard deviation
of $\alpha$ in our 20 events. For each event, we computed $\alpha$ values by applying Equation \ref{eqn:alpha_by_pix} to pixels in our flare masks where the field strength was greater than 300 G.  These masks should delineate the fields that participated in each flare's reconnection, and that might therefore be expected to exhibit more uniform post-flare twist.
Two standard deviations were separately calculated, one each from the sets of positive and negative $\alpha$ values. The standard deviation for each event is chosen to be the average of them. In Figure \ref{fig:twistVariancewithmask}, we show the difference in standard deviations in $\alpha$ versus the average of standard deviations.   This difference is negative for most of our flares 
(13 $<$ 0 versus 7 $>$ 0), 
meaning $\alpha$ tends to become more uniform in our flare sample.   
If positive and negative changes were equally likely, the probability that 13 or more events among our sample of 20 would have negative changes is about 13\% 
-- unlikely, but still a reasonable possibility.  The evolution of twist in our AR sample due to flares, therefore, tends to accord with expectations based on the Taylor relaxation hypothesis, but this effect is not strong.  
We also note that Equation 11 can be also written, from Equation (\ref{eqn:alphaav}), as 
\begin{equation}
\alpha_w = \frac{\displaystyle \int_{S} J_z(x, y) \, B_z(x, y) \, dx \,dy}{\displaystyle \int_{S} \, B_z^2(x, y)\, dx \, dy},
\end{equation}
To the extent that $B_z^2$ does not change significantly due to a flare, an increase of $\alpha_w$ indicates a larger $J_z$, albeit with an inverse-$|B_z|$ weighting.  

\subsection{Decomposition of Relative Helicity and Eruptive Events}
\label{ss:HjandHtotal}

\begin{figure}[ht]
\centering
\includegraphics[width=0.450\textwidth]{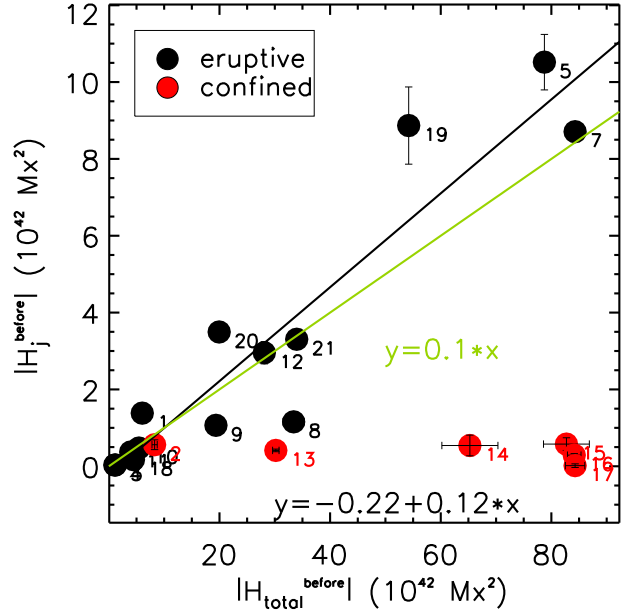}
\vspace{0.5cm}

\caption
{ Scatter plot between $|H_{total}|$ and $|H_j|$ 
for the events listed in Table \ref{table:xflares}. Black dots represent eruptive events; red dots represent confined events. Black line is a linear fit to the eruptive events (black dots). The green line, with a slope of 0.1, has been suggested by \citet{gupta2021} as a threshold that divides eruptive from confined events. Both $|H_{total}|$ and $|H_j|$ of AR 12192 for flares 13-17 are multiplied by a factor of 0.1 in order to be included in the plots. The events are denoted by their flare IDs.
}
\label{fig:ratioHjvsH}
\end{figure}

The helicity of the current-carrying field, $H_j$, has been found to be related to the eruptivity of events \citep[e.g.,][]{pariat2017}. The ratio of $|H_j|/|H|$ shows the potential to differentiate eruptive and non-eruptive events \citep[e.g.,][]{thalmann2019, gupta2021}. \citet{gupta2021} find a threshold value of 0.1 for this ratio. Above this threshold, flares will tend to be eruptive. Our results do not contradict this finding, as shown in Figure \ref{fig:ratioHjvsH}, albeit with statistical deviations. On the other hand, confined events clearly do not obey this scaling.  
Thus, as discussed in \citet{gupta2021}, the helicity ratio -- alone or in combination with other quantities such as the free energy ratio and flux-normalized current-carrying helicity -- may be useful to characterize the potential of an AR to flare.

\section{Discussion: Summary, Conclusions, and Future Work}
\label{s:conclusion}

To understand how magnetic energy and magnetic helicity evolve in association to major flares, we analyzed a dataset of photospheric vector magnetograms observed by HMI around the times of 21 GOES X-class flares produced by 13 active regions. Six of these flares, from two active regions, were confined; the remaining 15 flares, from 11 active regions, were eruptive. 

We applied the DAVE4VM tracking method to these magnetogram sequences to estimate the photospheric velocities that are needed to determine the fluxes of magnetic energy (the Poynting flux) and magnetic helicity across the photosphere. We also employed an optimization-based NLFFF code to extrapolate to coronal magnetic fields.  For these model coronal fields, we calculated the magnetic energy, free magnetic energy, relative magnetic helicity, and helicity of the current-carrying component of the field.  From the magnetograms themselves, we also calculated average twist over each AR (both unweighted and with $B_z^2$-weighting), and total unsigned flux in strong-field ($|{\bf B}| > 300$ G) pixels.   

To synthesize information from all events in order to characterize magnetic evolution associated with these flares, we used superposed epoch (SPE) analysis to combine time series from all regions.  For each of the quantities above, we extracted time series in a time interval around each event, and averaged these series with the flares' start times (the ``key times'' in SPE terminology) co-aligned.   We applied SPE to intervals of two durations:  within $\pm$ 5 hours of each flare, to identify typical rapid changes related to, or resulting from, flares; and, separately, from 48 hours before to 18 hours after each flare, to identify patterns in longer-time-scale evolution both leading up to flares and in the hours following them.

Here, we itemize our most notable results:

\begin{itemize}

\item Our main finding is that, on short time scales ($\sim$ 1h) after {\em eruptive} flares we find decreases in both the  coronal magnetic energy, $E$, and helicity, $H$ (Figure \ref{fig:speHandE}); and their photospheric fluxes, $dE/dt$ and $dH/dt$ (Figure \ref{fig:spe4ehflux}).  Changes in free energy are statistically significant; changes in total energy are less significant, given the roughly unchanged bulk energy of the reference, potential magnetic field. 

\item Using SPE to combine helicity estimates from our coronal extrapolations, we find clear evidence of decreases in helicity in eruptive flares (Figures \ref{fig:speHandE} and \ref{fig:changeHE4eruptiveflares}), consistent with the bodily removal of helicity from the corona by jettisoned coronal ejecta. \citet{Low2001} argued that the systematic removal of magnetic helicity from the corona by CMEs might play an important role in the solar dynamo. 
The insignificant decreases of total helicity that we find with confined flares suggests these flares are an insignificant sink for helicity, in accordance with the helicity conservation principle in confined systems. 

\item Looking at the evolution of energy and helicity over longer time scales (Figure \ref{fig:SPE4H_evo}) after eruptive events, we find that both quantities return to near-pre-flare levels.  The time scale for this replenishment is roughly 12 hours for the X-class flares analyzed in our work.  If flares substantially deplete the free energy in ARs, this would imply that there should be a relative paucity of multiple X-class flares from the same ARs on time scales shorter than this.  We find, however, that eruptive flares only remove 10\% of free energy (Figure \ref{fig:speHandE}, middle row).
Replenishment of helicity and energy after large eruptive events is consistent with the idea, suggested by \citet{Longcope2000}, that the solar interior can act as a twist reservoir (or ``current driver''), to re-supply coronal helicity as it is removed by ejections. 

\item Consistent with the previous report of \cite{Bobra2016}, we find that the twist present in our small sample of ARs that produced confined flares is roughly an order of magnitude smaller than the twist in our sample of ARs that produced eruptions (see Figure \ref{fig:spe4twist4twomethods}).

\item  Prior to eruptions, we find a tendency for average measures of twist within flare masks to decrease.  This might be explained by a toy-model with an emerging flux rope with large internal twist, surrounded by lower-twist areas: in the first phase of emergence, the photosphere cuts through the rope's apex, and a large area with high twist would be present in magnetograms; but as the rope's legs start to emerge, the area with strong, nonzero twist would decrease; sufficient emergence could trigger eruption.  We note that Figure \ref{fig:SPE4Mflux} shows a small but significant, pre-eruption tendency for total unsigned flux to increase, a signature of new flux emergence.  Our averages of $\alpha$ might also decrease if the pre-eruptive structure gradually ascends in a pre-flare, slow-rise process, which might decrease horizontal field strengths at the photosphere, lowering averaged values of $\alpha$.

\item  Following eruptions, 
average magnetic twist in ARs is also observed to increase significantly, as well as virtually instantly (i.e., within an hour or so - Figure \ref{fig:spe4twist4twomethods}). The causes of this increase in twist remain unclear.  
This might be explained by the strengthening of horizontal fields after eruptions, due to implosion of fields beneath the coronal reconnection region: increases in $|\alpha|$ would be expected if low-lying fields in eruptive regions have more organized twist (i.e., flux ropes) than the structures hosting confined flares. By studying the standard deviation of strong-field pixels' twist before and after individual flares, we also find some evidence for Taylor relaxation (i.e., twist growing more uniform due to magnetic reconnection): there is a marginally significant preponderance of ARs with decreases in the standard deviation of twist after flares.  

\item Flare-associated changes in $dE/dt$ and $dH/dt$ are shown in Figure \ref{fig:spe4ehflux}.  The physical causes of the fluxes’ decrease is unclear, although we see clear evidence that properties of the photospheric flows that drive these fluxes change after flares.   Average flow speeds in our event sample do not change significantly.  Delving deeper into flow changes, we computed the pre- and post-flare distributions of pixels' horizontal and vertical velocities as functions of those pixels' total field strength (Figure \ref{fig:densitydifferencemap}).  We see increases in speeds among weak-field pixels and decreases in speeds among strong-field pixels.  These changes might yield decreases in the magnitudes of $dE/dt$ and $dH/dt$.  It is known that strong flares often cause the magnetic field in some areas to become ``more horizontal'' due to an increase in $B_h$, with the normal field relatively unchanged  \citep[e.g.,][]{Hudson2008,Wang2010,Barczynski2019}. Given that stronger fields more effectively suppress convection (e.g., \citealt{Berger1998, Welsch2012}), flare-associated increases in $B_h$ might be partly responsible for decreases in speeds in high-$|{\bf B}|$ pixels.  Such changes, however, typically occur over small areas, meaning that, by themselves, magnetic changes might be a minor factor in altering the overall fluxes.  We also computed the pre- and post-flare distributions of pixels' Poynting flux values as functions of those pixels' horizontal and vertical magnetic field, and, separately, as functions of their horizontal and vertical velocities (Figure \ref{fig:densitydiff4energy}). Studying the differences between the pre- and post-flare distributions, we see clear increases in downward Poynting fluxes after flares.  The nature of these post-flare increases in the downward flow of magnetic energy warrants further study.   	

\item For confined flares, only changes in the free magnetic energy are found to be statistically significant. Given our small sample of confined flares, however,  we cannot be confident that a larger sample of such events would not exhibit at least some of the changes we see in our eruptive sample.  

\end{itemize}

In addition to the physical implications of our results for an improved understanding of the roles of energy and helicity in strong flares, we briefly refer to two methodology implications here: 
first, the SPE approach is useful for finding patterns that could well be missed in studies of individual events, because: (i) for an individual AR, time series of the variables that we study exhibit strong fluctuations; and (ii) there is substantial variation between ARs.
Second, our methods of estimating both the photospheric fluxes of energy and helicity and their coronal values are robust enough to enable detection of systematic changes in these quantities, despite fluctuations in each AR's evolution and variability between ARs. The importance of combining diagnostics from multiple, independent sources when magnetic energy and helicity are studied was also highlighted by \citet{Thalmann2021}. 
  
Our results suggest several areas for future research:
\begin{itemize}

\item Our analysis of magnetic evolution associated with confined flares was severely hampered by our small sample size.  A similar analysis on a larger sample, particularly for confined events, would be both worthy and meaningful. 

\item The nature of increases in downward Poynting flux is not yet understood.  In which magnetic structures do such changes occur?  What effect(s) cause(s) these changes?   One approach on addressing these questions is to undertake case studies, and in each case to identify areas with increases in downward Poynting flux after flares.  Magnetic structures in these areas, and their relationships to flare morphology, such as ribbon emission, should be investigated.

\item The nature of pre-flare decreases and post-flare increases in AR twist associated with large, eruptive flares is also yet to be understood.  From differences in pre-flare and post-flare magnetograms, where do the primary changes in twist occur?  What is the magnetic structure in areas with significant changes in twist?  Are all such areas closely associated with the flare site, or does twist sometimes change in significant ways in areas peripheral to the core flare location? Again, case studies might be valuable.  

\end{itemize}

We believe our results leave a lot to be understood and be learned about flares and eruptions from studies of photospheric vector magnetograms, along with physical quantities such as photospheric velocity fields and extrapolatd coronal magnetic fields, to name a few, that are derived from them. 


\acknowledgments  We wish to thank the anonymous referee for the valuable comments and suggestions. BTW, SHP, and YL thank the US taxpayers for providing the funding that made this research possible. We acknowledge support from NASA LWS 80NSSC19K0072 (YL and BTW).
NASA's SDO satellite and the HMI instrument were joint efforts by many teams and individuals, whose efforts to produce the HMI magnetograms
that we analyzed here are greatly appreciated. YG is supported by NSFC (11773016 and 11961131002) and 2020YFC2201201.
GV acknowledges funding by the Bundesministerium für Wirtschaft und Technologie through Deutsches Zentrum für Luft- und Raumfahrt e.V. (DLR), Grants No. 50 OT 1001/1201/1901 as well as 50 OT 0801/1003/1203/1703, and by the President of the Max Planck Society (MPG). EP acknowledges support from the French Programme National PNST of CNRS/INSU co-funded by CNES and CEA. EP also acknowledges financial support from the French national space agency (CNES) through the APR program.
JT acknowledges the Austrian Science Fund (FWF): P31413-N27. This article profited from discussions during the meetings of the ISSI International Team ``Magnetic Helicity estimations in models and observations of the solar magnetic field''
(http://www.issibern.ch/teams/magnetichelicity/) and the ISSI International Team ``Magnetic Helicity in Astrophysical Plasmas'' (https://www.issibern.ch/teams/helicityastroplas/).
BTW acknowledges support from the Fundaci\'on Jes\'us Serra, which supported a visit to the IAC-Tenerife during which a portion of this work was completed.

\bibliographystyle{aasjournal}
\bibliography{abbrevs,short_abbrevs,biblio,full_lib,bib_mods,biblio_mkg}


\end{document}